\documentclass[pre,superscriptaddress,showpacs]{revtex4}

\usepackage{graphicx}%
\usepackage{dcolumn}
\usepackage{amsmath}

\usepackage[dvips]{epsfig}
\usepackage{amssymb}
\usepackage{array,longtable}
\usepackage{amscd}

\begin{document}



\newcommand{\be}{\begin{eqnarray}}
\newcommand{\ee}{\end{eqnarray}}
\newcommand{\bse}{\begin{subequations}}
\newcommand{\ese}{\end{subequations}}

\newcommand{\bs}{\boldsymbol}
\newcommand{\mbb}{\mathbb}
\newcommand{\mcal}{\mathcal}
\newcommand{\mfr}{\mathfrak}
\newcommand{\mrm}{\mathrm}

\newcommand{\ovl}{\overline}
\newcommand{\p}{\partial}
\newcommand{\f}{\frac}
\newcommand{\diff}{\mrm{d}}
\newcommand{\lan}{\langle}
\newcommand{\ran}{\rangle}

\newcommand{\ga}{\alpha}
\newcommand{\gb}{\beta}
\newcommand{\gc}{\gamma}
\newcommand{\gd}{\delta}
\newcommand{\Gc}{\Gamma}
\newcommand{\gl}{\lambda}
\newcommand{\gk}{\kappa}
\newcommand{\go}{\omega}
\newcommand{\Go}{\Omega}
\newcommand{\veps}{\varepsilon}

\newcommand{\sn}{\mrm{sn}}
\newcommand{\cn}{\mrm{cn}}
\newcommand{\dn}{\mrm{dn}}
\newcommand{\am}{\mrm{am}}
\newcommand{\sech}{\mrm{sech}}
\newcommand{\sign}{\mrm{sign}}

\newcommand{\csp}{\;,\qquad\qquad}
\newcommand{\fa}{\forall\;}

\newcommand{\N}{\mbb{N}}
\newcommand{\R}{\mbb{R}}
\newcommand{\D}{\mcal{D}}
\newcommand{\Nn}{\mcal{N}}

\newcommand{\im}{\mrm{image}\;}
\newcommand{\num}{\mrm{\#}}


\title{Theory of Relativistic Brownian Motion: The (1+3)-Dimensional Case}

\author{J\"orn Dunkel}
\email{dunkel@mpa-garching.mpg.de}
\affiliation{Max-Planck-Institute for Astrophysics,
Karl-Schwarzschild-Stra{\ss}e 1, Postfach 1317,
D-85741 Garching, Germany}

\author{Peter H\"anggi}
\affiliation{Institut f\"ur Physik, Universit\"at Augsburg,
 Theoretische Physik I,  Universit\"atstra{\ss}e 1, D-86135 Augsburg, Germany}

\date{\today}

\begin{abstract}
A theory for (1+3)-dimensional relativistic Brownian
motion under the influence of external force fields is put forward. Starting out from
a set of  relativistically covariant, but multiplicative Langevin equations we describe the
relativistic stochastic dynamics of a forced Brownian particle.
The corresponding Fokker-Planck
equations are studied in the laboratory frame coordinates. In particular,
the stochastic integration prescription, i.e. the discretization rule dilemma,
is elucidated (pre-point discretization rule {\it vs.} mid-point discretization rule {\it vs.}
post-point discretization rule). Remarkably, within our relativistic scheme we find that
the post-point rule (or the transport form) yields the only
Fokker-Planck dynamics from which the relativistic
Maxwell-Boltzmann statistics is recovered as the stationary solution.
The relativistic velocity effects become distinctly more pronounced by going from one to three spatial dimensions.
Moreover, we present numerical results
for the asymptotic mean square displacement of a free relativistic
Brownian particle moving in (1+3) dimensions.
\end{abstract}

\pacs{
02.50.Ey, 
05.40.-a, 
05.40.Jc, 
47.75.+f  
}

\maketitle

\section{Introduction}
\label{introduction}

The problem of formulating a consistent theory of Brownian motions
\cite{Ei05c,EiSm,UhOr30,Ch43,WaUh45,KaSh91} in the framework of special
relativity \cite{Ei05a,Ei05b} represents a longstanding issue in mathematical
and statistical physics (classical references are \cite{Sc61,Ha65,Du65}; more
recent contributions include
\cite{GuRu78,Bo79a,BY81,MoVi95,Po97,OrHo03,FrLJ04,DeMaRi97,De04}; for
a kinetic theory approach, see \cite{St71,DG80}). In a preceding paper
\cite{DuHa04a} -- referred to as paper I hereafter -- we have discussed in
detail how one can construct Langevin equations for
(1+1)-dimensional relativistic Brownian motions. In particular, it was
demonstrated that for the relativistic Langevin equation {\it per se}
cannot uniquely determine the corresponding Fokker-Planck equation
(FPE). This ambiguity arises due to the fact that the relativistic Langevin
equations, when e.g. written in laboratory coordinates,
exhibit a multiplicative coupling between a function of the momentum
coordinate and a Gaussian white noise process (laboratory frame
$\equiv$ rest frame of the heat bath). Thus, depending on the choice
of the discretization rule, different forms of relativistic FPE are
obtained \cite{VK03,HaTo82}.
\par
In paper I, we have analyzed the three most popular discretization
rules for Langevin equations with multiplicative noise, which can be traced back to the
proposals pioneered by Ito (pre-point discretization rule \cite{Ito44,Ito51}), by
Fisk and Stratonovich (mid-point rule
\cite{St64,St66,Fisk63,Fisk65,HaTo82}), and by H\"anggi and
Klimontovich (post-point rule \cite{Han78,Han80,Han82,Kl94}). In this
context it could be shown that only the H\"anggi-Klimontovich (HK)
interpretation
of the Langevin equation yields the transport form of the Fokker-Planck equation
with state-dependent diffusion, whose stationary
solution coincides with the one-dimensional relativistic Maxwell
distribution. The latter is  known from J\"uttner's early work on the relativistic
gas \cite{Ju11,Sy57} and also from the relativistic kinetic theory
\cite{DG80}.
\par
In paper I, we have focussed  exclusively on the simplest situation, corresponding to
{\it free} (1+1)-dimensional relativistic Brownian motions. Therefore, the present
paper aims to extend the analysis to the physically relevant  (1+3)-dimensional
case. In
particular, we wish to  include as well  the effects of additional, {\it external}
force fields. To this end the paper is structured as follows: In
Sec. \ref{langevin_approach} the relativistic Langevin equations are
given in covariant 4-vector notation and also in laboratory frame
coordinates.  The corresponding Fokker-Planck equations and their
stationary solutions are considered in Sec. \ref{FPE}. Section
\ref{numerics} contains a discussion of numerical results for the mean
square displacement of free Brownian particles. The paper concludes
with a resume of the main results in Sec. \ref{summary}.

\section{Relativistic Langevin dynamics}
\label{langevin_approach}

Let us first discuss the manifestly Lorentz-covariant 4-vector form of the relativistic
Langevin equations (Sec. \ref{relativistic_generalization}). For that purpose,
we shall use of the results derived in Sec. II of paper I, which
can readily be generalized to (1+3)-dimensions. Subsequently, the relativistic
Langevin equations will be written in laboratory coordinates (Sec. \ref{lab}). The
latter form provides the basis for the numerical results of
Sec. \ref{numerics}.
\par
With regard to notation, the following conventions will be used throughout the paper: Upper and lower
Greek indices $\ga,\gb,\ldots$ can take values $0,1,2,3$, where \lq$0$\rq\space
refers to the time component. Upper and lower Latin indices $i,j,\ldots$ take values $1,2,3$
and are used to label the components of spatial 3-vectors,
denoted by bold symbols. For example, we write $(x^\ga)=(x^0,\bs
x)=(ct,x^i)$ and $(p^\ga)=(p^0,\bs p)=(E/c,p^i)$ with $t$ denoting the coordinate time, $E$ the
energy, $c$ the vacuum speed of light, $x^i$ and $p^i$ the spatial
coordinates and relativistic
momenta, respectively. Moreover, Einstein's summation convention is applied
throughout. The (1+3)-dimensional Minkowski metric tensor with respect to Cartesian coordinates is taken as
$$
(\eta_{\ga\gb})=(\eta^{\ga\gb})=\mrm{diag}(-1,1,1,1)\csp
({\eta^\ga}_\gb)=({\eta_\ga}^\gb)=\mrm{diag}(1,1,1,1).
$$
As commonly known, covariant vector components $x_\ga$ can be calculated from the
contravariant components $x^\ga$ by virtue of $x_\ga=\eta_{\ga\gb}x^\gb$,
which, in particular, means that for Cartesian coordinates $x_0=-x^0$
and $x_i=x^i$ holds. Further, if in a certain inertial
coordinate system $\Sigma$ a Brownian particle has the 3-velocity  $\bs
v(t)\equiv \diff \bs x(t)/\diff t$, then the differential $\diff \tau$ of its proper-time is defined by
\be
\diff \tau\equiv
\diff t \sqrt{1-\f{v_i v^i}{c^2}}.
\ee

\subsection{Langevin equations in 4-vector notation}
\label{relativistic_generalization}

Consider a Brownian particle with rest mass $m$, proper time $\tau$ and
4-velocity $u^\gb(\tau)$, i.e., the 4-momentum of the particle is given by
$p^\ga=m u^\ga$, where $u_\ga u^\ga \equiv -c^2$. Assume that the particle is surrounded by an isotropic,
homogeneous heat bath with constant 4-velocity $U^\gb$ and, additionally,
subject to an external 4-force $K^\ga(x^\nu,p^\mu)$ such as, e.g., the
Lorentz force. Then, according to the
results in Sec. II of paper I, the relativistic Langevin equations of
motion read
\bse\label{e:RBM_covariant}
\be
\diff x^\ga(\tau)&=&\label{e:RBM_covariant.1}
\f{p^\ga}{m} \diff \tau\\
\diff p^\ga(\tau)&=&\label{e:RBM_covariant.2}
\left\{K^\ga -{\nu^\ga}_\gb
\left[p^\gb-m U^\gb\right]\right\} \diff \tau
+w^\ga(\tau).
\ee
For an isotropic homogeneous heat bath, the friction tensor ${\nu^\ga}_\gb$
in \eqref{e:RBM_covariant.2}  is given by
\be\label{e:RBM_covariant.3}
{{\nu}^\ga}_\gb=
\nu\left({\eta^\ga}_{\gb}+\f{u^\ga u_\gb}{c^2}\right),
\ee
with $\nu$ denoting the scalar viscous friction coefficient measured in the
rest frame of the particle. Furthermore, the relativistic Wiener increments
$w^\ga(\tau)\equiv \diff W^\ga(\tau)$ are distributed according to the
probability  density
\be
\mcal{P}^{1+3}[w^\ga(\tau)]
=\label{e:RBM_covariant_dens}
\f{c}{(4\pi D\; \diff \tau)^{3/2}}\;
\exp\left[-\f{w_\ga(\tau)\,w^\ga(\tau)}{4D\,\diff \tau}\;\,\right]
\times \gd\left[u_\ga\,w^\ga(\tau)\right],
\ee
\ese
where $D$ is the scalar noise amplitude parameter measured in the
rest frame of the particle. Some useful comments concerning Eq. \eqref{e:RBM_covariant} are appropriate:
\par
(i) The  covariant friction tensor in Eq. \eqref{e:RBM_covariant.3} carries the same structure as
the covariant pressure tensor for an ideal fluid. In particular, this means
that in each instantaneous rest frame $\Sigma_*$ of the particle, where
temporarily $(u_*^\ga)=(c,\bs 0)$ holds, the tensor  form Eq. \eqref{e:RBM_covariant.3} reduces to the diagonal
form
\be\label{e:friction_tensor}
({{\nu_*}^\ga}_\gb)=\begin{pmatrix}
0&0&0&0\\
0&\nu&0&0\\
0&0&\nu&0\\
0&0&0&\nu
\end{pmatrix}.
\ee
This form of the friction tensor reflects the simplifying assumptions that the
heat bath can, in good approximation, be considered as isotropic and
homogeneous.
\par
(ii) Note that  the probability
density of the increments in Eq. \eqref{e:RBM_covariant_dens} can equivalently be written as
\bse
\be\label{e:RBM_covariant_dens-2}
\mcal{P}^{1+3}[w^\ga]
=
\f{c}{(4\pi D\; \diff \tau)^{3/2}}\;
\exp\left[-\f{1}{2\diff\tau}\hat{D}_{\ga\gb}w^\ga\,w^\gb\right]
\times \gd\left[u_\ga\,w^\ga\right],
\ee
where the tensor
\be
{{\hat D}^\ga}_{\;\gb}=
\f{1}{2D}\left({\eta^\ga}_{\gb}+\f{u^\ga u_\gb}{c^2}\right)
\ee
\ese
carries the same isotropic  structure as the friction tensor from
Eq. \eqref{e:RBM_covariant.3}. The $\delta$-function in
Eqs. \eqref{e:RBM_covariant_dens-2} accounts for the fact that the
Minkowski scalar-product of the 4-force and the 4-velocity must
identically vanish.
\par
(iii) The increment density in Eq. \eqref{e:RBM_covariant_dens} is
normalized so that
\bse\label{e:properties}
\be
1=
\left\{\prod_{\ga=0}^3\int_{-\infty}^\infty \diff[w^\ga(\tau)]\right\}
\mcal{P}^{1+3}[w^\ga(\tau)]
\ee
holds. For the first two moments one finds
\be
\lan w^\ga(\tau)\ran & =&0,\label{e:RBM_covariant.4} \\
\lan w^\ga(\tau)\, w^\gb(\tau') \ran &=&\label{e:RBM_covariant.5}
\begin{cases}
0, & \tau\ne \tau';\\
{D}^{\ga\gb} \diff \tau,& \tau= \tau',
\end{cases}
\ee
where
\be
{{D}^\ga}_\gb=
2D \left({\eta^\ga}_{\gb}+\f{u^\ga u_\gb}{c^2}\right).
\ee
\ese
The easiest way to validate this is to perform the calculations leading to
Eqs. \eqref{e:properties} in a co-moving Lorentz frame $\Sigma_*$, where, at a
given instant of time $t_*(\tau)$, the particle is at rest. In such a
co-moving frame $\Sigma_*$ the marginal distribution of the spatial momentum increments, defined by
\be\label{e:marginal}
\mcal{P}^{3}[\bs w_*(t_*)]=\int_{-\infty}^\infty \diff w^0_*\;
\mcal{P}^{1+3}[w_*^\ga],
\ee
reduces to a Gaussian. One thus recovers, as it should hold true,  from Eqs. \eqref{e:RBM_covariant} the
nonrelativistic Brownian motion in the Newtonian limit case $\bs v^2\ll c^2$.

\subsection{Langevin dynamics in the laboratory frame}
\label{lab}

In this section the covariant Langevin equations
\eqref{e:RBM_covariant} will be rewritten in laboratory coordinates. A laboratory frame $\Sigma_0$ is,
by definition, an inertial system, in which
the heat bath is at rest. That is, in $\Sigma_0$ the 4-velocity of the heat
bath is given by $(U^\gb)=(c,\bs 0)$ for all times $t$, where $t$ is
the coordinate time of $\Sigma_0$.
\par
From Eq. \eqref{e:RBM_covariant.1}, we obtain three differential
equations for the position coordinates
\bse\label{e:lab-EOM}
\be\label{e:lab-EOM-c}
\diff x^i(t)= v^i\, \diff t,
\ee
where
\be\label{e:v-p-identity}
v^i=
\f{c p^i}{\sqrt{m^2c^2+ p_ip^i}}.
\ee
Furthermore, since we have $(U^\gb)=(c,\bs 0)$ in $\Sigma_0$, the four stochastic
differential Eqs. \eqref{e:RBM_covariant.2} can be rewritten as
\cite{DuHa04a,Weinberg}
\be
\diff {p}^i&=&\label{e:lab-EOM-a}
(\gc^{-1}K^i-\nu\, {p}^i)\, \diff t+ w^i,\\
\diff E &=&\label{e:lab-EOM-b}
(\gc^{-1} K^i-\nu\, p^i) v_i\, {\diff t}+ c w^0,
\ee
where the relativistic (kinetic) energy is here defined by $E\equiv c p^0$, and
\be \label{e:gamma}
\gc
\equiv
\left(1-\f{v_i v^i}{c^2}\right)^{-1/2}
=
\left(1+\f{p_i p^i}{m^2c^2}\right)^{1/2}.
\ee
\ese
Before discussing the stochastic increments $w^\ga$, let us briefly
consider the deterministic force components $K^i$. If $\bs F=(F^i)$ is the nonrelativistic (Newtonian) force, then the corresponding relativistic force 3-vector $\bs K= (K^i)$ is given by
(see Chap. 2 of \cite{Weinberg})
\be\label{e:force}
\bs K= \bs F +(\gc-1)\f{(\bs v \cdot \bs F)}{v^2}\; \bs v.
\ee
If the dynamics is  confined to one spatial dimension only, then Eq. \eqref{e:force}
simplifies to $\bs K=\gc \bs F$.
\par
Next, let us turn to the stochastic force components, appearing on the rhs. of
Eqs. \eqref{e:lab-EOM-a} and \eqref{e:lab-EOM-b}. According to
Eq. \eqref{e:RBM_covariant_dens}, the  distribution of the stochastic
momentum increments $w^\ga$, also depends on the particle's velocity
$\bs v$. By virtue of the relation
\be
(u_\ga)=\left(-\gc c,\gc v_i\right),
\qquad\qquad
(w^\ga)=\left(w^0, w^i\right),
\ee
we can rewrite the increment density \eqref{e:RBM_covariant_dens} in
laboratory coordinates as follows
\be\label{e:lab_probability_density}
\mcal{P}^{1+3}[w^\ga]=
c\left(\f{\gc}{4\pi D\; \diff t}\right)^{3/2}
\exp\left[-\f{w_i w^i-(w^0)^2}{4D\;\diff t/\gc}
\right]
\times \gd\left(c\gc w^0-\gc v_i w^i \right).
\ee
As we already pointed out above, the $\delta$-function in Eq.
\eqref{e:lab_probability_density} reflects the fact that the energy increment
$w^0$  is coupled to the spatial momentum increments $w^i$ via
\be
0=u_\ga w^\ga=-c\gc w^0+\gc  v_i  w^i
\qquad\Rightarrow\qquad
w^0=\f{v_i w^i}{c}.
\ee
This is just the stochastic analogue of the well-known deterministic
identity $$u_\ga K^\ga\equiv 0. $$ Hence, similar to $K^0$, also $w^0$
can be eliminated from the Langevin equation \eqref{e:lab-EOM-b}, yielding
\be
\diff E
=
(\gc^{-1}K^i-\nu\, p^i)v_i\, {\diff t}+  v_i  w^i(t)
=\label{e:lab-EOM-1.2}
 v_i\,\diff p^i.
\ee
Using the identity \eqref{e:v-p-identity}, we can further rewrite Eq. \eqref{e:lab-EOM-1.2} as
\be
\diff E=\f{cp_i}{\sqrt{m^2c^2+ \bs{p}^2c^2}}\,\diff {p^i},
\ee
where $\bs{p}^2\equiv p_i p^i$. From the preceding equation we regain
the well-known energy momentum law
\be\label{e:energy}
E(t)=\sqrt{m^2c^4+\bs{p}(t)^2c^2}.
\ee
It might be worthwhile to remark that in the presence of force
fields the relativistic energy $E=c p^0$ is generally different from the
relativistic Hamiltonian \cite{Go80}. Note that Eq. \eqref{e:energy}
remains valid also for our stochastic model.
\par
The relativistic Brownian motion is therefore completely described by the
three Langevin equations \eqref{e:lab-EOM-a}.  The statistics of the increments $w^i$ in
\eqref{e:lab-EOM-a} is determined by the marginal distribution
$\mcal{P}^{3}[\bs w]$, defined in
Eq. \eqref{e:marginal}. Performing the integration over the
$\delta$-function in Eq. \eqref{e:lab_probability_density}, we find
\be
\mcal{P}^{3}[\bs w]
&=&\notag
\int_{-\infty}^\infty \diff w^0\;\mcal{P}^{1+3}[w^\ga]\\
&=&\notag
c\int_{-\infty}^\infty \diff w^0\;
\left(\f{\gc}{4\pi D\; \diff t}\right)^{3/2}
\exp\left[-\f{w_i w^i-(w^0)^2}{4D\;\diff t/\gc}
\right]
\times \gd\left(c\gc w^0-\gc v_i w^i \right)\\
&=& \label{e:lab_density}
\f{1}{\gc}\left(\f{\gc}{4\pi D \diff t}\right)^{3/2} \exp\left[
-\f{\gc}{4D \diff t}\left(\gd_{ij}-\f{v_i v_j}{c^2}\right)
w^i w^j \right],
\ee
where $\gd_{ij}$ denotes the Kronecker-symbol (defined by $\gd_{ij}=1$
if $i=j$, and $\gd_{ij}=0$ otherwise).
\par
In principle, it is straightforward  to perform computer simulations on the basis of Eqs. \eqref{e:lab-EOM-a} and
\eqref{e:lab_density}.  In Sec. \ref{numerics} we will discuss several
numerical findings. Before doing so, however, it is worthwhile to consider in
greater detail the Fokker-Planck equations of the relativistic Brownian
motion in the laboratory frame $\Sigma_0$. By doing so it will become clear that the choice $v^i=v^i(t)$ in
Eq. \eqref{e:lab_density} is consistent with an Ito-interpretation
\cite{Ito44,Ito51,VK03} of the stochastic differential equations
\eqref{e:lab-EOM-a}. However, we shall also see that alternative
interpretations -- such as e.g. the post-point discretization rule, $v^i=v^i(t+dt)$ -- lead to physically reliable
results as well.

\section{Relativistic Fokker-Planck equations}
\label{FPE}

The objective here is to discuss relativistic Fokker-Planck
equations (FPE) for the one-particle momentum density $f(t,\bs p)$  and, as well,
for the phase space density $f(t,\bs p,\bs x)$. In the remainder,
we will exclusively refer to the coordinates of the laboratory frame
$\Sigma_0$. Before turning to the relativistic FPE in
Sects. \ref{FPE-r-free} and \ref{FPE-r-lorentz}, it is useful to briefly recall the nonrelativistic
case.

\subsection{Nonrelativistic case}
\label{FPE-nr}

Consider the nonrelativistic Langevin equations
\bse\label{e:FPE-nr-LE}
\be \label{e:FPE-nr-LE-a}
\diff x^i  &=& v^i \,\diff t\\
\diff p^i  &=& (K^i-\nu\, p^i) \diff t + w^i,
\ee
where $p^i(t)=m v^i(t)$ denotes the nonrelativistic momentum components,
$K^i=-\p^i U$ represents the vector components of a conservative force with
time-independent potential $U(\bs x)$, and where the increments $w^i\equiv
\diff W^i$ are distributed according to
\be \label{e:FPE-nr-LE-b}
\mcal{P}[\bs w]=
\left(\f{1}{4\pi D\,\diff t}\right)^{3/2}
\exp\left[-\f{w_i\, w^i}{4D\,\diff t}\right].
\ee
\ese
Equations \eqref{e:FPE-nr-LE} govern the nonrelativistic motions of a Brownian particle in
the rest frame of the heat bath. As one readily observes, in the case of conservative force fields,
Eqs. \eqref{e:FPE-nr-LE} can be obtained from the relativistic equations
\eqref{e:lab-EOM-c}, \eqref{e:lab-EOM-a} and \eqref{e:lab_density}  by formally taking the limit $c\to \infty$ (Newtonian limit
case). It is well known that the phase space density $f(t,\bs p,\bs
x)$, associated with the stochastic process \eqref{e:FPE-nr-LE}, is governed by the FPE \cite{VK03,Schwabl,HaTo82}
\be\label{e:FPE-nr-f(p,x)}
\f{\p}{\p t} f+\f{p^i}{m}\f{\p}{\p x^i} f +\f{\p}{\p p^i}(K^i f)=
\f{\p}{\p p^i}\left(\nu p^i f + D \gd^{ij} \f{\p}{\p p^j} f\right).
\ee
The stationary solution of Eq. \eqref{e:FPE-nr-f(p,x)} is the (nonrelativistic)
Maxwell-Boltzmann distribution, i.e.,
\bse
\be\label{e:MB}
f(\bs p,\bs x)=
C \exp\left[-\f{\bs p^2 +2m\,U(\bs x)}{2m\, k_\mrm{B} T}\right],
\ee
where $C$ is a normalization constant. The temperature  $T$ of the
bath is defined by the Einstein relation ($k_\mrm{B}$ denotes the Boltzmann constant)
\be
k_\mrm{B} T\equiv\f{D}{m \nu}.
\ee
\ese
The related marginal momentum distribution is the usual Maxwellian probability density
\be\label{e:FPE-nr-sol}
f(\bs p)
=\left(\f{\nu}{2\pi D}\right)^{3/2} \exp\left(-\f{\nu \bs p^2}{2D}\right).
\ee

\subsection{Relativistic FPE for free Brownian particles}
\label{FPE-r-free}

We next reason three different types of relativistic Fokker-Planck
equations for the momentum density $f(t,\bs p)$ of a free Brownian particle,
whose dynamics is governed by the  stochastic process given by Eqs. \eqref{e:lab-EOM-a} and
\eqref{e:lab_density}. The corresponding equations for the phase-space
density $f(t,\bs p,\bs x)$ will be considered separately in Sec.
\ref{FPE-r-lorentz}, where we will also include the influence of an external force field.
\par
As our starting point serves the relativistic Langevin equations
\eqref{e:lab-EOM-a}, which hold in the laboratory frame $\Sigma_0$
(i.e., in the rest frame of the heat bath). In absence of external force fields
the stochastic differential equations \eqref{e:lab-EOM-a} reduce to
\bse\label{e:lab-EOM-w}
\be
\diff {p}^i&=&
-\nu\,p^i\, \diff t+w^i,
\ee
where $p^i=\gc mv^i$ is now the relativistic momentum and $\gc$ has been
defined in Eq. \eqref{e:gamma}. According to Eq. \eqref{e:lab_density} --
and in distinct contrast to Eq. \eqref{e:FPE-nr-LE-b} -- the distribution of
the relativistic increments  reads
\be
\mcal{P}^{3}[\bs w]=
\f{1}{\gc}\left(\f{\gc}{4\pi D \diff t}\right)^{3/2} \exp\left[
-\f{w_i\, {A^i}_j\, w^j}{4D \diff t} \right]
\ee
with matrix elements given by
\be
{A^i}_j
\equiv\left({\gd^i}_j-\f{v^i v_j}{c^2}\right)\gc
=\left({\gd^i}_j-\f{p^i p_j}{\gc^2m^2c^2}\right)\gc.
\ee
\ese
Following the reasoning of paper I, the next  aim is to rewrite the Langevin equations
\eqref{e:lab-EOM-w} in such a form, that the resulting equations
exhibit multiplicative Gaussian white noise, governed by a
velocity-independent normal
distribution of the form \eqref{e:FPE-nr-LE-b}.
In order to achieve this objective we first note that the matrix $\bs A(\bs
p)=({A^i}_j)$ is symmetric. Its eigenvalues and determinant are given by
\be\label{e:spectrum_A}
\mrm{spec}(\bs A)=\{\gc,\gc,\gc^{-1}\},\qquad
\det(\bs A)=\gc.
\ee
Thus, the matrix  $\bs A$ is positive definite for velocities  $\bs v^2
<c^2$, and the elements of the inverse matrix  $\bs A^{-1}$ read
\be\label{e:inverse_A}
{(A^{-1})^j}_k
=\left(\f{{\gd^j}_k}{\gc^2}+\f{v^j v_k}{c^2}\right)\gc
=\left({\gd^j}_k+\f{p^j p_k}{m^2c^2}\right)\f{1}{\gc}.
\ee
Furthermore, there exists a unique Cholesky-decomposition \cite{Te1}
\be
\bs A=\bs L^\top \bs L=
\begin{pmatrix}
{L^1}_1 & 0      & 0  \\
{L^2}_1 & {L^2}_2 & 0  \\
{L^3}_1 & {L^3}_2 & {L^3}_3
\end{pmatrix}
\begin{pmatrix}
{L^1}_1 & {L^2}_1 & {L^3}_1 \\
 0      & {L^2}_2 & {L^3}_2 \\
 0      & 0       & {L^3}_3
\end{pmatrix},
\ee
where the matrix $\bs L(\bs p)$ is {\em nonsingular} with elements given by
\be
{L^1}_1&=&\sqrt{{A^1}_1},\\
{L^2}_1&=&{A^2}_1/{L^1}_1,\notag\\
{L^2}_2&=&\sqrt{{A^2}_2-({L^2}_1)^2},\notag\\
{L^3}_1&=&{A^3}_1/{L^1}_1,\notag\\
{L^3}_2&=&({A^3}_2- {L^3}_1 {L^2}_1)/{L^2}_2,\notag\\
{L^3}_3&=&\sqrt{{A^3}_3- ({L^3}_1)^2-({L^3}_2)^2}.\notag
\ee
The inverse matrix $\bs L(\bs p)^{-1}$ reads
\bse\label{e:inverse_L}
\be
\bs L^{-1}=\f{1}{\det(\bs L)}
\begin{pmatrix}
{L^2}_2 {L^3}_3 & -{L^2}_1 {L^3}_3 &{L^2}_1{L^3}_2 -{L^3}_1{L^2}_2 \\
 0      & {L^3}_3  {L^1}_1&- {L^3}_2{L^1}_1 \\
 0      & 0       &{L^1}_1 {L^2}_2
\end{pmatrix},
\ee
where
\be
\det(\bs L)={L^1}_1{L^2}_2 {L^3}_3.
\ee
\ese
Let us next introduce a stochastic vector variable $\bs y(t)=(y^i(t))$ by
\be
y^i\equiv {L^i}_j w^j.
\ee
Then, by taking into account that for $\bs w^\top\equiv (w_i)$ and  $\bs y^\top\equiv (y_i)$ the relation
\be
\bs w^\top \bs A \bs w
=\bs w^\top \bs L^\top \bs L \bs w
=(\bs L\bs w)^\top \bs L \bs w
=\bs y^\top \bs y
\ee
holds, we can rewrite the Langevin equation \eqref{e:lab-EOM-w} in the form
\bse\label{e:lab-EOM-y}
\be
\diff \bs p&=&\label{e:lab-EOM-y-1} -\nu
\,\bs p\, \diff t+\bs{L}(\bs p)^{-1}\bs{y},
\ee
where $\bs y(t)$ is distributed according to the momentum-independent Gaussian probability
density
\be
\mcal{P}_y^3[\bs y] &=&\label{e:lab-EOM-y-2}
\left(\f{1}{4\pi D\; \diff t}\right)^{3/2}
\exp\left[ -\f{y_i\, y^i}{4D\;\diff t}\right].
\ee
\ese
Put differently, because the inverse matrix $\bs{L}(\bs p)^{-1}$ depends on the
momentum coordinate $\bs p$, the random vector $\bs y(t)$ enters in
relativistic Langevin equation \eqref{e:lab-EOM-y-1} as an ordinary \lq
multiplicative\rq\space  Gaussian white noise process with noise strength
$D$.
\par
As is well known, for multiplicative stochastic processes of the type  \eqref{e:lab-EOM-y}
the Langevin equation {\it per se} does
{\em not} uniquely determine a corresponding Fokker-Planck equation
\cite{HaTo82,VK03}. In the following subsections, we shall
discuss the three most popular choices of resulting Fokker-Planck equations
for a Langevin equation of the form \eqref{e:lab-EOM-y}. These choices are rooted in
the different proposals put forward by Ito
\cite{Ito44,Ito51,HaTo82,VK03}, by Stratonovich and Fisk
\cite{St64,St66,Fisk63,Fisk65,HaTo82}, and by H\"anggi
\cite{Han78,Han80,Han82} and Klimontovich \cite{Kl94}, respectively.
All three approaches have in common that the related Fokker-Planck
equation can be written as a continuity equation (conservation of probability) of the form
\cite{Han82}
\be\label{e:continuity}
\f{\p}{\p t} f(t, \bs p)+\f{\p}{\p
p^i} j^i(t, \bs p)=0,
\ee
but with distinct different expressions for the probability
current $\bs j(t, \bs p)$. It is worthwhile to anticipate here that only the
H\"anggi-Klimontovich approach (see Sec. \ref{FPE-jutt}) yields a
stationary distribution, which can be identified with  J\"uttner's
relativistic Maxwell distribution \cite{Ju11}.

\subsubsection{Ito approach}
\label{FPE-ito}

According to Ito's interpretation of the Langevin equation
\eqref{e:lab-EOM-y-1}, the coefficient matrix before $\bs y(t)$ is to be
evaluated at the lower boundary of the interval $[t,t+\diff t]$, i.e.,
\be\label{e:ito_choice}
\bs L(\bs p\bigr)^{-1}=\bs L\bigl(\bs p(t)\bigr)^{-1}.
\ee
Ito's choice is also known as the pre-point discretization rule \cite{Ito44,Ito51,HaTo82,VK03} and leads
to the following explicit expression for the current

\be\label{e:ito_current-a}
j_\mrm{I}^i(t,\bs p)=
-\left\{\nu p^i f+D\f{\p}{\p p_j}
{\left[\bs{L}^{-1}(\bs{L}^{-1})^\top\right]^i}_j\; f\right\}.
\ee
In view of the identity
\be
\bs{L}^{-1}(\bs{L}^{-1})^\top
=\bs{L}^{-1}(\bs{L}^\top)^{-1}
=(\bs{L}^\top\bs{L})^{-1}=
\bs A^{-1},
\ee
Eq. \eqref{e:ito_current-a} can be rewritten more conveniently as
\be\label{e:ito_current}
j_\mrm{I}^i(t,\bs p)=
-\left[\nu p^i f+
D\f{\p}{\p p_j} {(A^{-1})^i}_j f\right],
\ee
where the matrix $\bs A(\bs p)^{-1}$ is given in Eq. \eqref{e:inverse_A}.
The related relativistic Fokker-Planck equation is obtained by inserting
this current into the conservation law \eqref{e:continuity}. As elucidated
in the Appendix \ref{a:solution}, the stationary solution of the resulting Fokker-Planck equation reads
\be\label{e:ito_solution}
f_\mrm{I}(\bs p)=
C_\mrm{I}\left(1+\f{\bs p^2}{m^2c^2}\right)^{-3/2}
\exp\left(-\chi\sqrt{1+\f{\bs p^2}{m^2c^2}}\right),
\ee
which contains in the prefactor an explicit (non-Maxwell like) dependence on velocity
and where $C_\mrm{I}$ is a normalization constant. In the solution
\eqref{e:ito_solution}, the dimensionless
parameter
\be
\chi\equiv \f{\nu m^2c^2}{D}
\ee
is related to the scalar temperature $T$ of the heat bath via the
Einstein relation
\be\label{e:temperature}
k_\mrm{B} T\equiv \f{mc^2}{\chi}=\f{D}{m\nu}.
\ee
Thus, the quantity $\chi=mc^2/(k_\mrm{B}T)$ gives the ratio between
rest energy and thermal energy of the Brownian particle. It should be
mentioned that we used in paper I the notation $\gb$
instead of $\chi$. However, in order to avoid a possible confusion with the
commonly used abbreviation $\gb=(k_\mrm{B}T)^{-1}$ we opted
here for this slight change of notation.

\subsubsection{Stratonovich-Fisk approach}
\label{FPE-strat}

According to the Stratonovich-Fisk prescription, the coefficient matrix before $\bs y(t)$ in
 \eqref{e:lab-EOM-y-1} is to be evaluated with the mid-point discretization rule, i.e.,
\be
\bs L(\bs p)^{-1}
=
\bs{L}\biggl(\f{\bs p(t)+\bs p(t+\diff t)}{2}\biggr)^{-1}.
\ee
This choice \cite{St64,St66,Fisk63,HaTo82} leads to a different expression for the  current
\cite{St64,St66,Fisk63,HaTo82}:
\be\label{e:stratonovich_current}
j_\mrm{SF}^i(t,\bs p)=
-\left\{\nu p^i f+D\, {(L^{-1})^i}_k
\f{\p}{\p p_j}
{\left[(L^{-1})^\top\right]^k}_j  \; f\right\}.
\ee
The explicit stationary solution of Stratonovich's Fokker-Planck equation reads (see Appendix \ref{a:SF})
\be\label{e:stratonovich_solution}
f_\mrm{SF}(\bs p)=
C_\mrm{SF} \left(1+\f{\bs p^2}{m^2c^2}\right)^{-3/4}
\exp\left(-\chi \sqrt{1+\f{\bs p^2}{m^2c^2}}\right),
\ee
and thus differs from Eq. \eqref{e:ito_solution} in the power of the velocity-dependent prefactor.

\subsubsection{H\"anggi-Klimontovich approach}
\label{FPE-jutt}

Ultimately, let us next consider the H\"anggi-Klimontovich (HK) stochastic integral interpretation,
sometimes  referred to as the transport form
\cite{Han78,Han80,Han82} or also as the kinetic form \cite{Kl94}.
According to this interpretation, the coefficient matrix in front of $\bs y(t)$ in
\eqref{e:lab-EOM-y-1} is to be evaluated at the upper boundary of the
interval $[t,t+\diff t]$; i.e., within the post-point discretization we set
\be
\bs L(\bs p)^{-1}=\bs{L}\bigl(\bs p(t+\diff t)\bigr)^{-1}.
\ee
This choice leads to the following expression for the current
\cite{Han80,Han82,Kl94}
\be\label{e:K_current}
j_\mrm{HK}^i(t,\bs p)=-\left[\nu p^i f+D\; {(A^{-1})^i}_j \f{\p}{\p p_j} f\right],\ee
and the stationary solution of the related FPE reads (see Appendix Eq. (\ref{a:HK}))
\bse \label{e:juettner}
\be\label{e:K_solution}
f_\mrm{HK}(\bs p)=
C_\mrm{HK}\exp\left(-\chi \sqrt{1+\f{\bs p^2}{m^2c^2}}\right).
\ee
Note that this solution contains no velocity dependence in the prefactor.
Using the temperature definition in \eqref{e:temperature} and the relativistic
kinetic energy formula $E=\sqrt{m^2 c^4+\bs p^2c^2}$, we can recast
\eqref{e:K_solution} as
\be
f_\mrm{HK}(\bs p)= C_\mrm{HK}\exp\left(-\f{E}{k_\mrm{B}T}\right).
\ee
The normalization constant is given by \cite{Du65}
\be
C_\mrm{HK}^{-1}
=
\int \diff^3\bs p\; \exp\left(-\chi\,\sqrt{1+\f{\bs p^2}{m^2c^2}} \right)
=4\pi
\int_{0}^\infty \diff p\;p^2\, \exp\left(-\chi\,\sqrt{1+\f{p^2}{m^2c^2}} \right)
=4\pi (mc)^3 \f{K_2(\chi)}{\chi},
\ee
\ese
where $K_2(\chi)$ denotes the modified Bessel function of the second kind.
The distribution function \eqref{e:juettner} is known as the
{\it relativistic Maxwell distribution}. It was first derived by F. J\"uttner
\cite{Ju11} in 1911, when he investigated the velocity distribution of
non-interacting relativistic gas particles (see also \cite{Sy57}). By comparing
\eqref{e:ito_solution}, \eqref{e:stratonovich_solution} and
\eqref{e:K_solution} one readily observes that the stationary solutions
$f_\mrm{I/SF}$ differ from the J\"uttner function $f_\mrm{HK}$ through
additional $\bs p$-dependent prefactors. The quantitative difference between
these three stationary solutions becomes significant in the relativistic
limit, corresponding to a low rest energy-to-temperature ratio $\chi$.
\par
As already mentioned in paper I, for the one-dimensional case the relativistic
Maxwell distribution has also been obtained by Schay, see Eq. (3.63) and
(3.64) in Ref. \cite{Sc61}, who studied relativistic diffusions employing
a transfer probability method.  Moreover, the distribution
\eqref{e:juettner} can also be derived in the framework of the relativistic
kinetic theory \cite{DG80}. This suggests that the HK-discretization rule is physically
preferable, if one wishes to use the above Langevin equations for numerical
simulations of relativistic kinetic processes. In general, however, additional
information about the microscopic structure of the heat bath is required, in
order to decide which discretization rule is physically reasonable (see, e.g.,
the discussion of Ito-Stratonovich dilemma in the context of \lq
internal/external\rq\space noise as given in Chap. IX.5 of van Kampen's
textbook \cite{VK03}).
\par
Finally, we still note that the related velocity probability density functions
$\phi_\mrm{I/SF/HK}(\bs v)$ are obtained by applying the general transformation law
\be
\phi(\bs v)
\equiv \label{e:velocity_density}
f(\bs p(\bs v))\left|\f{\p \bs p}{\p \bs v}\right|,
\ee
where, as usual,
$$
\bs p(\bs v)= \f{m \bs v}{\sqrt{1-\bs v^2/c^2}}.
$$
The determinant factor
\be
\left|\f{\p \bs p}{\p \bs v}\right|
=m^3 \left(1-\f{\bs v^2}{c^2}\right)^{-5/2},
\ee
appearing on the rhs. of  Eq. \eqref{e:velocity_density},
ensures that the velocity density functions $\phi_\mrm{I/SF/HK}(\bs v)$
drop to zero for $\bs v^2 \to c^2$. For example, in the case of the
J\"uttner function \eqref{e:juettner} one  explicitly obtains
\be
\phi_\mrm{HK}(\bs v) =
\f{\chi}{4\pi c^3K_2(\chi)}
 \exp\left(-\f{\chi}{\sqrt{1-{\bs v^2}/{c^2}}}
\right) \left(1-\f{\bs v^2}{c^2}\right)^{-5/2}.
\ee

\subsection{The inclusion of external force fields}
\label{FPE-r-lorentz}

The preceding section concentrated on Fokker-Planck equations for the momentum
density $f(t,\bs p)$. In this part we shall  discuss the corresponding resulting
equations for the one-particle phase space density $f(t,\bs p,\bs x)$.
As before, we refer to the coordinates of the laboratory frame
$\Sigma_0$, in which the heat bath is at rest.
\par
If an external force field is present, then  Eq. \eqref{e:lab-EOM-y-1}
generalizes to
\be
\diff \bs p&=&
(\gc^{-1} \bs K-\nu \,\bs p)\, \diff t+\bs{L}(\bs p)^{-1}\bs{y},
\ee
where $\bs y$ is distributed according to the momentum-independent Gaussian
density  from Eq. \eqref{e:lab-EOM-y-2}. The related relativistic
Fokker-Planck equation for the full phase space density $f(t,\bs p,\bs
x)$ thus reads
\be\label{e:FPE-rel-ex1}
\f{\p}{\p t}f+\f{p^i}{m \gc} \f{\p}{\p x^i} f
+\f{\p}{\p p^i} \biggl(\f{K^i}{\gc} f\biggr)
= - \f{\p}{\p p^i} j^i_\mrm{I/SF/HK}
\ee
where the current densities $\bs j_\mrm{I/SF/HK}(t,\bs p,\bs x)$ are
obtained by replacing  $f(t,\bs p)$ with $f(t,\bs p,\bs x)$ in the above
expressions for $\bs j_\mrm{I/SF/HK}(t,\bs p)$, respectively. In the limiting case that
$\nu\to 0$, $D\to 0$, the rhs. of Eq. \eqref{e:FPE-rel-ex1} becomes
equal to zero and one regains the relativistic Liousville equation or,
equivalently, the collision-less Boltzmann-Vlasov equation \cite{Liboff90,DG80}.\\
\par
As a particular example, let us consider a relativistic Brownian
particle with rest charge $q$, being subject to a static
electromagnetic field $(\bs E,\bs B)$, measured with respect to
$\Sigma_0$. Then, in addition to the stochastic interaction
with the heat bath, the deterministic Lorentz-force \cite{MiThWe00}
\be
\f{\bs K(\bs p,\bs x)}{\gc} =
q\left[\bs E(\bs x) + \f{\bs p}{m \gc c} \times \bs B(\bs x)\right]
\ee
is acting on the particle, where \lq$\times$\rq\space denotes the
exterior vector product. For simplicity, let us confine ourselves to
the HK-form of the FPE and let $\bs E(\bs x)=\nabla \Phi(\bs x)$ and
$\bs B(\bs x)\equiv \bs 0$ in the laboratory system. In this case, the
stationary solution of Eq. \eqref{e:FPE-rel-ex1} emerges as
\bse\label{e:MB-rel}
\be
f_\mrm{HK}(\bs p, \bs x)&=&
C_\mrm{HK} \exp\biggl\{-\chi\left[\gc(\bs p) - \f{q\Phi(\bs x)}{m c^2}\right]\biggr\}\\
&=&C_\mrm{HK}
\exp\biggl[-\f{E(\bs p) - q\Phi(\bs x)}{k_\mrm{B} T}\biggr],
\ee
\ese
where $C_\mrm{HK}$ is a normalization constant, and $E(\bs
p)=(m^2c^4+\bs p^2 c^2)^{1/2}$ denotes the relativistic (kinetic)
energy. It is reassuring to see that the solution \eqref{e:MB-rel} just represents the
relativistic generalization of the nonrelativistic Maxwell-Boltzmann
distribution from Eq. \eqref{e:MB}.

\section{Numerical investigations}
\label{numerics}

The numerical results presented in this section were obtained on the basis of
the relativistic Langevin equations \eqref{e:lab-EOM-y}, which hold in the
laboratory frame $\Sigma_0$. For simplicity, we confined ourselves to
considering free Brownian particles (i.e. $\bs K=\bs 0$) and employed the
Ito-discretization scheme with fixed time step $\diff t$, see Sec. \ref{FPE-ito}.
In all simulation we have used an ensemble size
of $N=1000$ particles. A characteristic unit system was fixed by
setting $m=c=\nu=1$. Formally, this corresponds to considering re-scaled
dimensionless quantities, such as $\tilde{p}^i=p^i/(mc)$, $\tilde{x}^i=x^i
\nu/c$, $\tilde{t}=t\nu$, $\tilde{v}^i=v^i/c$  etc.. The simulation time-step
was always chosen as $\diff t =0.001 \nu^{-1}$, and the Gaussian random
variables $y^i(t)$ were generated with the pseudo-random number
generator of Mathematica \cite{Ma03}.

\subsection{Distribution functions}
\label{distributions}

\begin{figure}[h]
\center
\epsfig{file=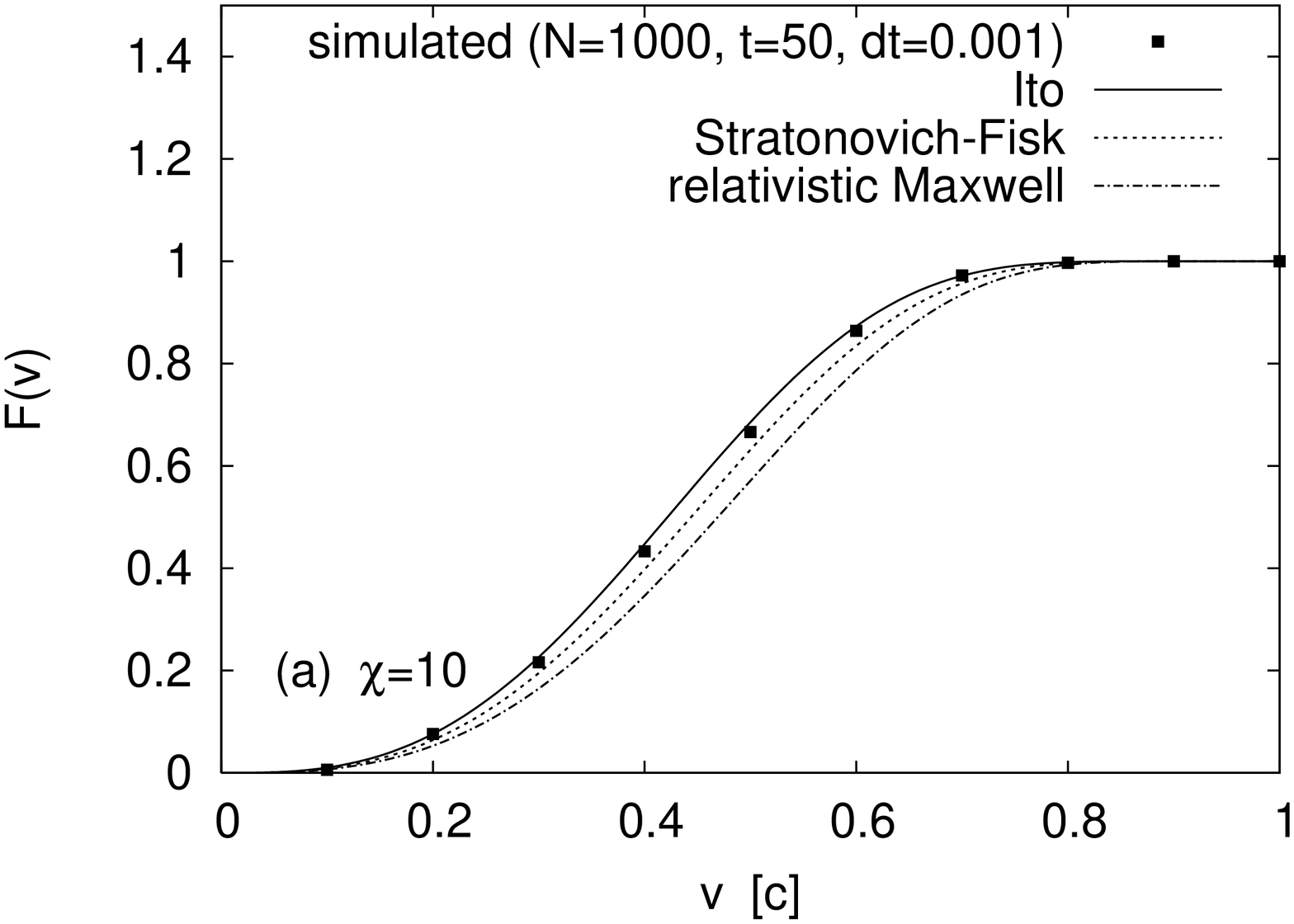,height=7.9cm, angle=-90}
\epsfig{file=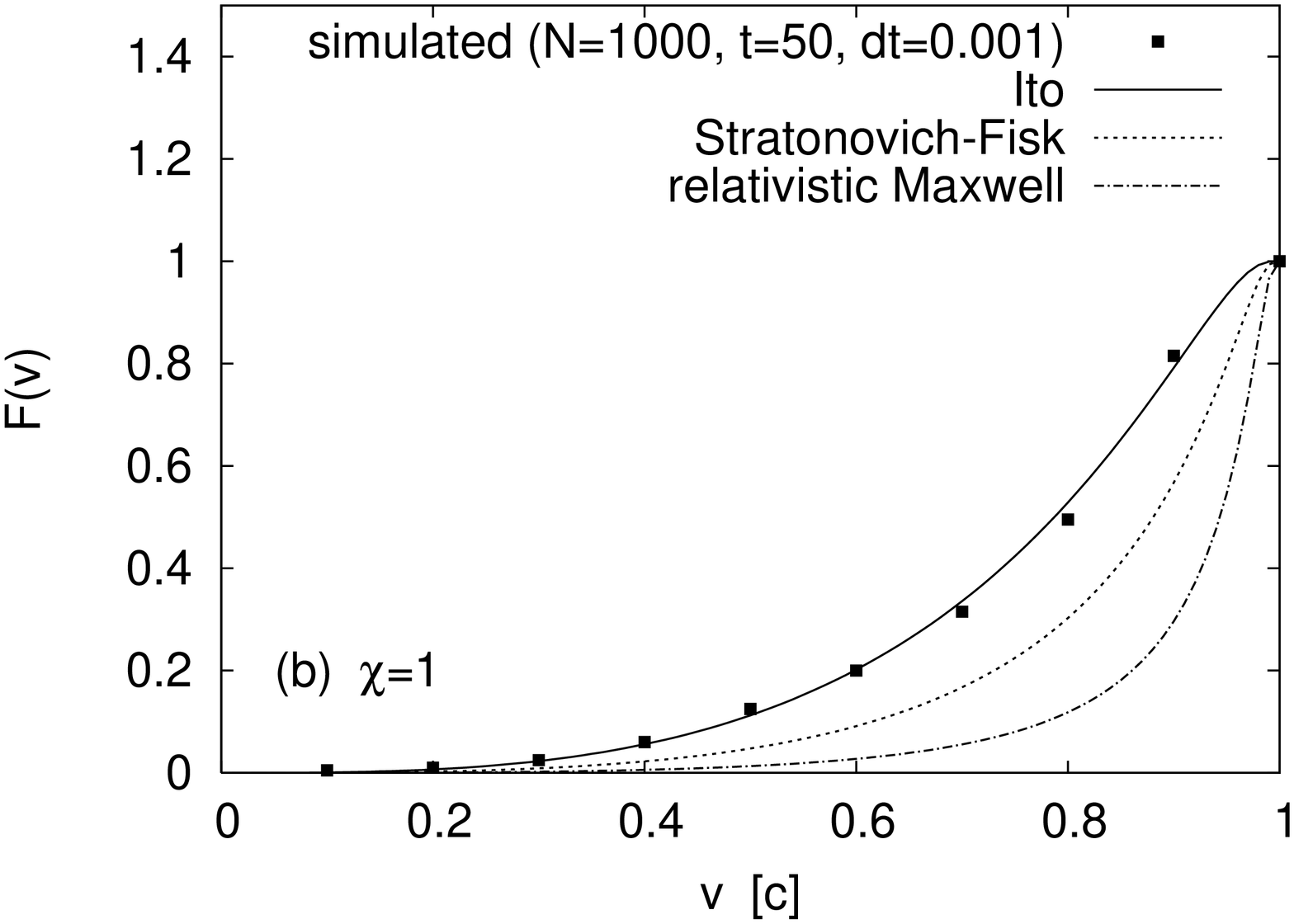,height=7.9cm, angle=-90}
\caption{These diagrams depict a comparison among the numerical Ito-prrescription
 and the three analytical
  results for the corresponding stationary cumulative distribution function $F(v)$ in the laboratory frame
  $\Sigma_0$. (a) In the nonrelativistic limit $\chi\gg 1$ the stationary
  solutions of the three different FPE are nearly indistinguishable. (b) In the strong
   relativistic limit case $\chi \le 1$, however, the stationary solutions exhibit
  deviations from each other. Because our simulations are based on an
  Ito-discretization  scheme, the numerical data points do agree
  with the Ito solution (solid line). \label{fig01}}
\end{figure}

In the simulations we have numerically measured the stationary cumulative
distribution function $F$ of the absolute velocity values $v\equiv \sqrt{v_i v^i}$
in the laboratory frame $\Sigma_0$. Given, e.g., a spherically symmetric
probability density $\phi(\bs v)\equiv \tilde\phi(v)$ with normalization
\be
1=
\int \diff^3 \bs v\;\phi(\bs v)
=
4\pi \int_{0}^1\diff v\;v^2\tilde \phi(v),
\ee
the respective cumulative distribution function is defined by
\be\label{e:distribution_function}
F(v)= 4\pi\,\int_{0}^v\diff u\;u^2\;\tilde \phi( u).
\ee
In order to obtain $F(v)$ from numerical simulations, one simply measures
the relative fraction of particles with absolute velocities in the interval
$[0,v)$. Figure \ref{fig01} depicts the numerically
determined  stationary distribution functions, taken at
time $t=100 \nu^{-1}$  and also the corresponding analytical curves
$F_\mrm{I/SF/HK}(v)$. The latter were obtained by numerically integrating the
formula \eqref{e:distribution_function}, using the three different stationary
density functions $\phi_\mrm{I/SF/HK}(\bs v)=\tilde\phi_\mrm{I/SF/HK}(v)$ found in Sec. \ref{FPE}.
\par
As one can see in diagram \ref{fig01} (a), for low temperature values,
corresponding to  $\chi\gg 1$, the three stationary distribution
functions approach each other, since they all converge to the
nonrelativistic Maxwell distribution in the limit $\chi\to \infty$. For high
temperatures, corresponding to $\chi \le 1$, the stationary solutions
exhibit significant quantitative differences, cf. Fig. \ref{fig01}
(b). Since our simulations are based on an Ito-discretization  scheme,
the numerical data points agree best  with the Ito solution (solid
line). Similar to the (1+1)-dimensional case \cite{DuHa04a}, the
quality of the fit remains satisfactory over several orders of
magnitudes of the parameter $\chi$. One may therefore conclude that the numerical simulations of Langevin
equations provide a useful tool for studying relativistic Brownian
motions in (1+3) dimensions. It should, however,  be stressed again, that the
appropriate choice of the discretization rule is particularly
important with regard to potential applications to physical situations.
\par
\begin{figure}[h]
\center
\epsfig{file=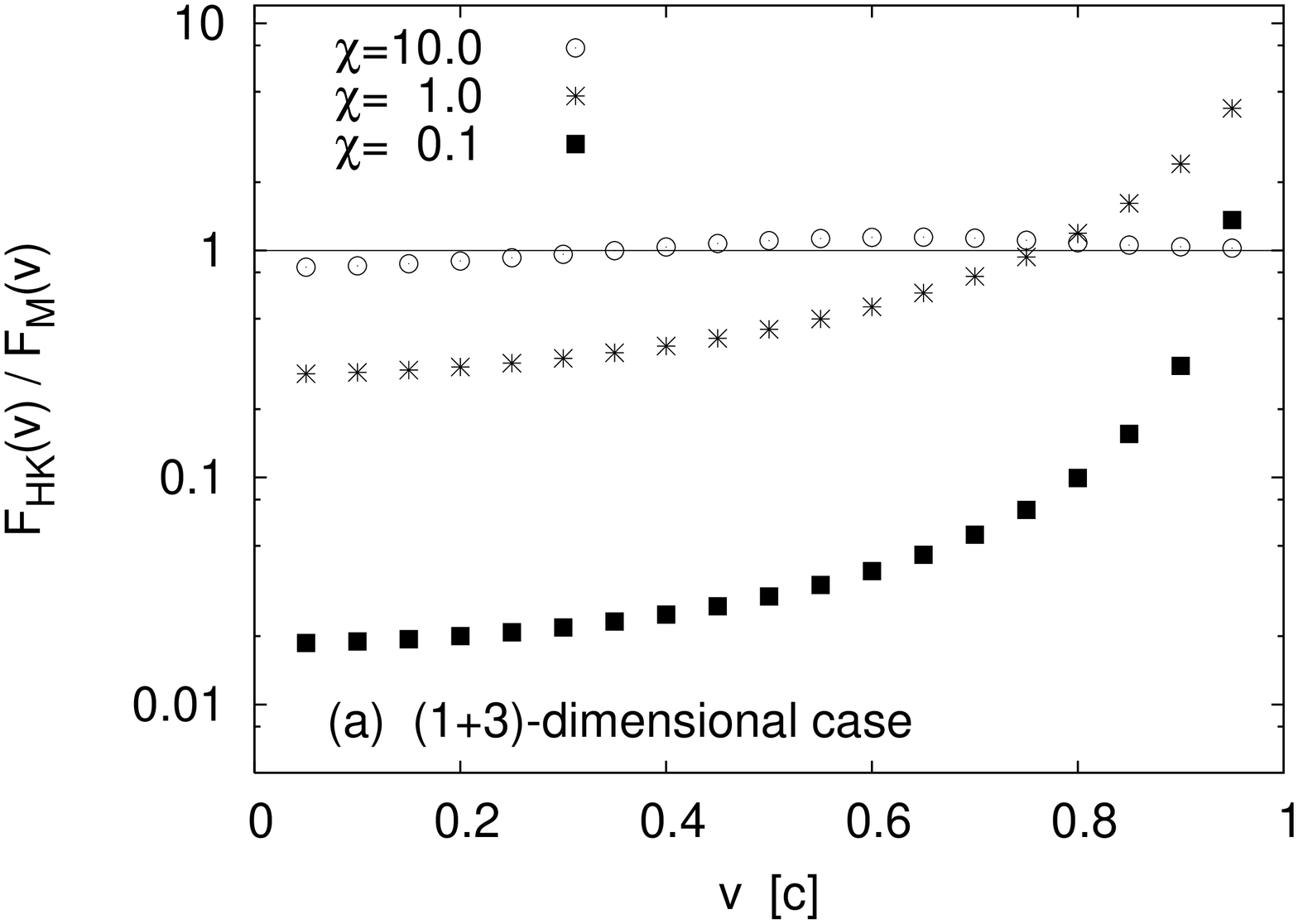,height=8cm, angle=-90}
\epsfig{file=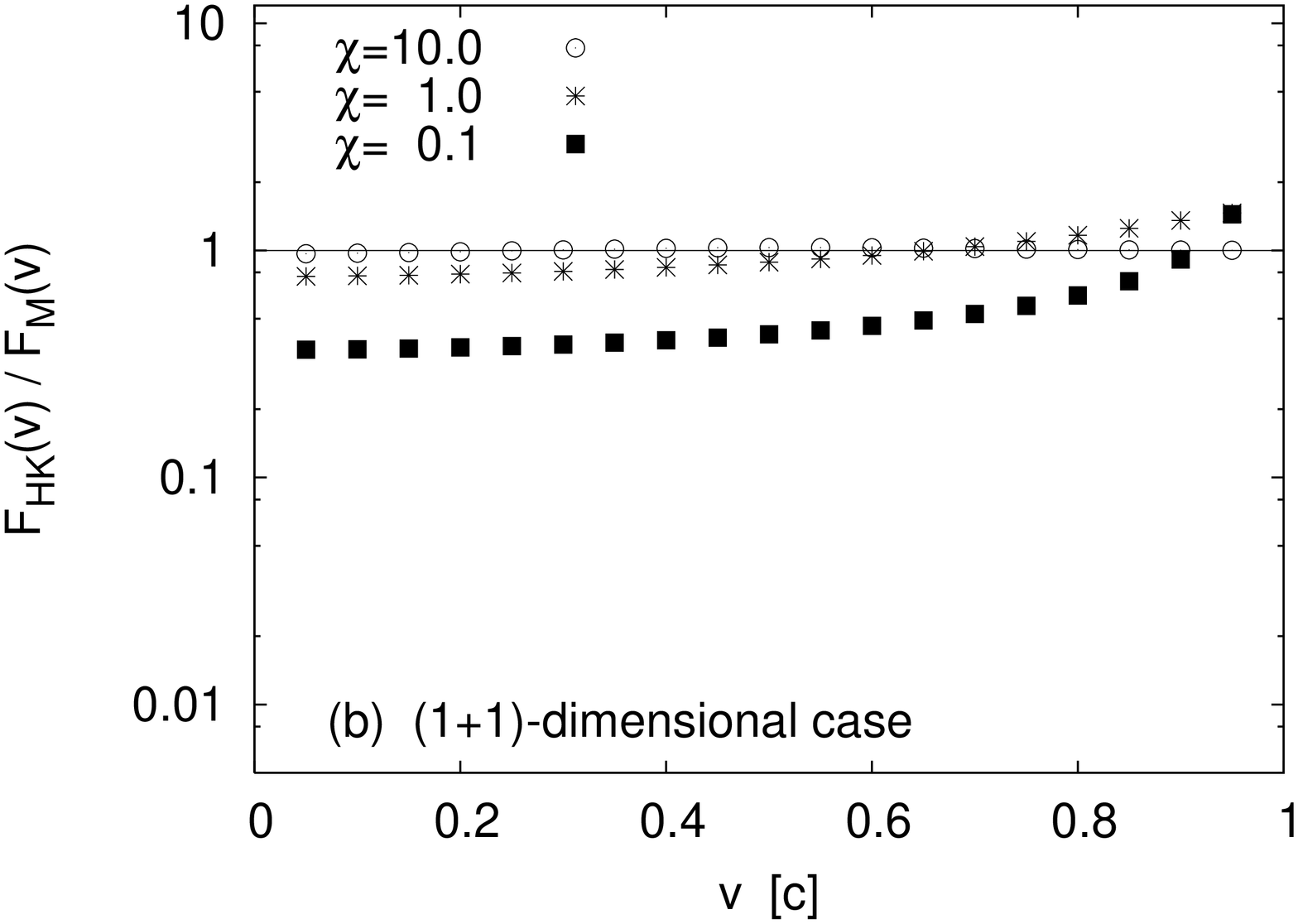,height=8cm, angle=-90}
\caption{The diagrams show the ratio $F_\mrm{HK}(v)/F_\mrm{M}(v)$
for three different values of the characteristic parameter
$\chi=mc^2/(k_\mrm{B}T)$, where $F_\mrm{M}$ and $F_\mrm{HK}$ denote the cumulative
distribution functions of absolute velocity for the nonrelativistic and
relativistic Maxwell distributions, respectively. Note that
 the deviation between the nonrelativistic and the
corresponding relativistic distribution functions are more pronounced (at fixed value of
 $\chi$) for the (1+3)-dimensional situation, i.e.
in Fig. (a). Therefore, compared with the (1+1)-dimensional case in Fig. (b),
relativistic effects become detectable best at finite, lower temperatures, if the Brownian dynamics
proceeds in (1+3)-dimensions.
\label{fig02}}
\end{figure}
Furthermore, Fig. \ref{fig02} illustrates the influence of the number of spatial
dimensions on the occurrence of relativistic effects. In the two
diagrams we depicted the ratio $F_\mrm{HK}(v)/F_\mrm{M}(v)$
for three different values of the characteristic parameter $\chi$, with
$F_\mrm{HK/M}(v)$ denoting the cumulative velocity distribution function of the
relativistic and nonrelativistic Maxwell distribution, respectively.
Figure \ref{fig02} (a) corresponds to the case of (1+3)-dimensional free
Brownian motions, while Fig. \ref{fig02} (b) refers to the
(1+1)-dimensional case, discussed in paper I. The comparison of the two
diagrams reveals that relativistic effects become
significantly enhanced at  lower temperatures (or larger values of $\chi$,
respectively), if the Brownian particle moves in
(1+3)-dimensions.

\subsection{Mean square displacement}
\label{mean_square_displacement}

Next, the spatial mean square displacement of the free relativistic Brownian motion
is investigated. Since this quantity is easily accessible in
experiments, it has played an important role in the verification of the
nonrelativistic theory.
\par
As before, we consider an ensemble of $N$ independent, free Brownian particles
with coordinates $\bs x_{(i)}(t)$ in $\Sigma_0$ and
initial conditions $\bs x_{(i)}(0)=\bs 0, \bs v_{(i)}(0)=\bs 0$ for $i=1,2,\ldots, N$.
The position mean value is obtained as
\be
\ovl{\bs x}(t)\equiv \f{1}{N}\sum_{i=1}^N \bs x_{(i)}(t),
\ee
and the related second moment is given by
\be
\ovl{\bs x^2}(t)\equiv \f{1}{N}\sum_{i=1}^N \left[\bs x_{(i)}(t)\right]^2.
\ee
The empirical mean square displacement can then be defined as follows
\be
\sigma^2(t)\equiv \ovl{\bs x^2}(t)-\left[ \ovl{\bs x}(t)\right]^2.
\ee
Important results of the nonrelativistic theory of the three-dimensional
Brownian motion read
\bse
\be
\lim_{t\to +\infty} \ovl{\bs x}(t)&\to&\label{e:mean-a}\bs 0,\\
\lim_{t\to +\infty} \f{\sigma^2(t)}{t}&\to& 3\cdot 2D^x,
\ee
\ese
where the constant
\be\label{e:diffusion}
D^x=\f{k_\mrm{B} T}{m\nu}=\f{D}{m^2\nu^2}
\ee
is the nonrelativistic coefficient of diffusion in coordinate space [not to be
confused with noise strength  $D=kT/(m\nu)$].
\par
\begin{figure}[h]
\center
\epsfig{file=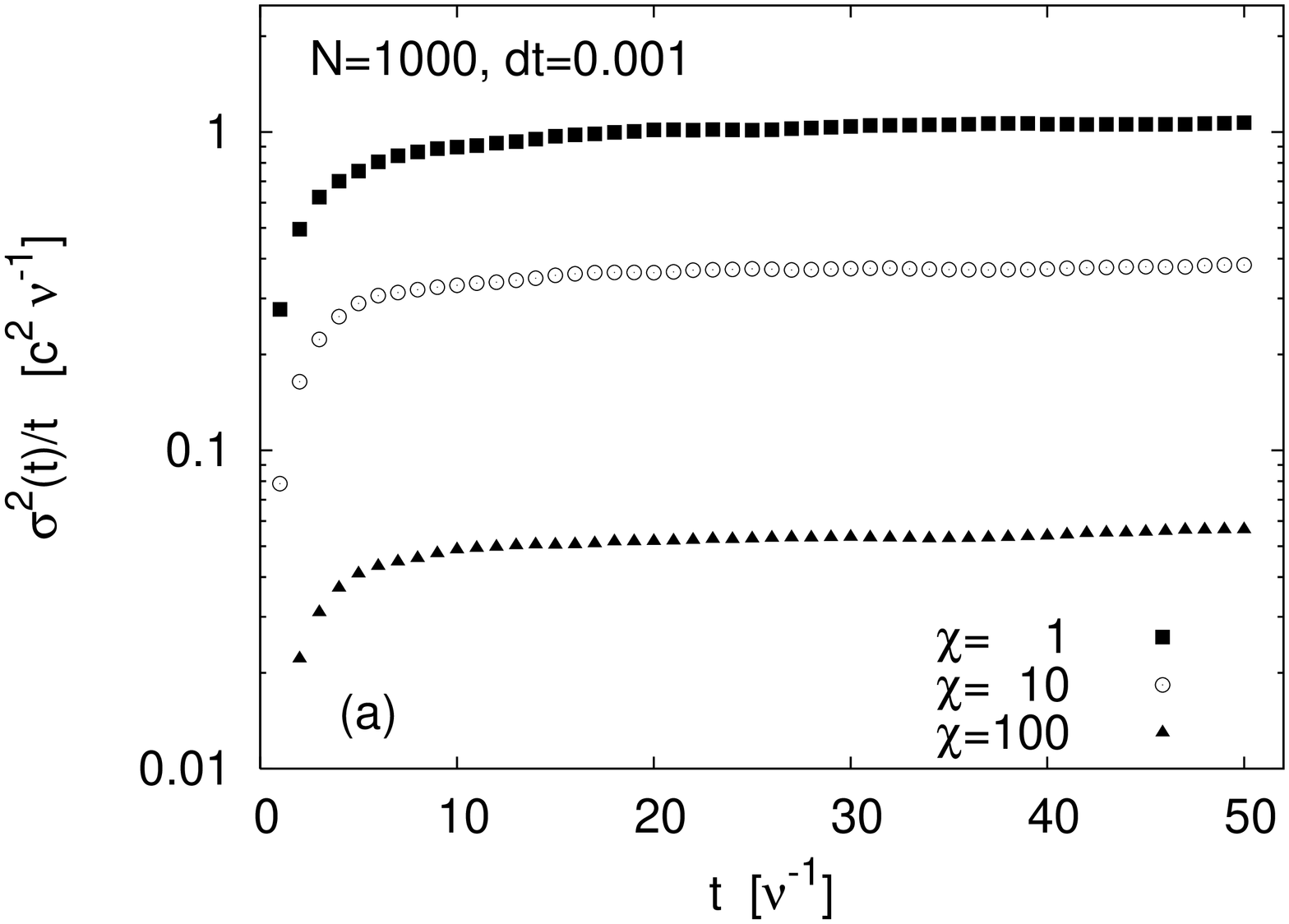,height=8cm, angle=-90}
\epsfig{file=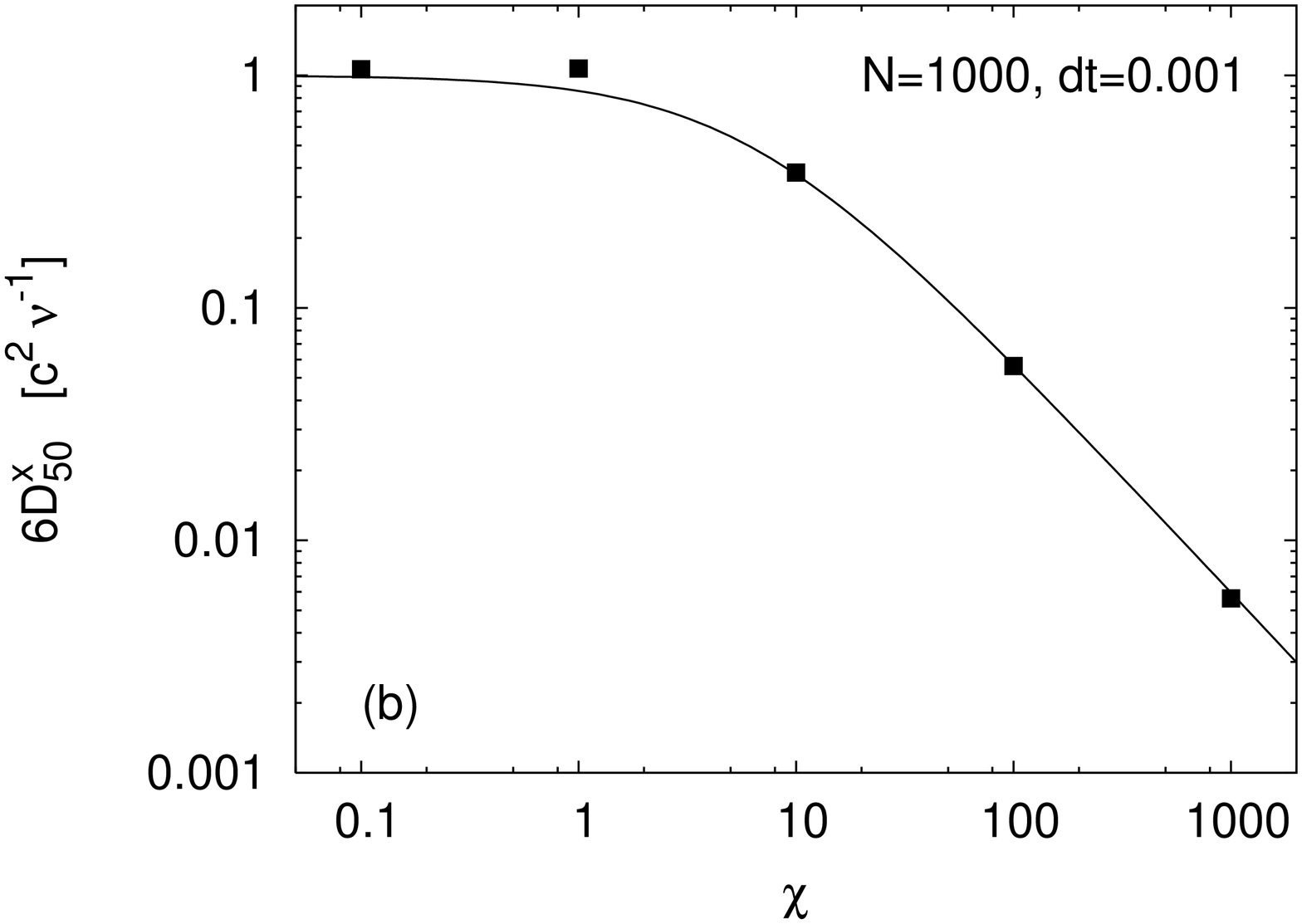,height=8cm, angle=-90}
\caption{(a) Mean square displacement, divided by elapsed time $t$, as numerically
  calculated for different $\chi$-values in the laboratory frame
  $\Sigma_0$ (rest frame of the heat bath). As evident from this diagram, for
  the relativistic Brownian motion the related asymptotic mean square
  displacement grows linearly with $t$. (b)  The coordinate space diffusion
  constant  $D^x_{50}(\chi)$ was numerically determined at time $t=50
  \nu^{-1}$. The solid line corresponds to the empirical fitting formula
  $D^x(\chi)=c^2\nu^{-1}(\chi+6)^{-1}$, which reduces to the
  classical nonrelativistic result  $D^x\simeq  c^2/(\nu\chi) =kT/(m\nu)$ for $\chi \gg 6$.
\label{fig03}}
\end{figure}

It is therefore interesting to consider
the asymptotic behavior of the quantity $\sigma^2(t)/t$ for relativistic
Brownian motions. In Fig. \ref{fig03} (a) one can see the corresponding
numerical results for different values of $\chi$, evaluated on the basis
  of the Ito-scheme from Sec. \ref{FPE-ito}.
As evident from this diagram, for each value of $\chi$, the quantity $\sigma^2(t)/t$ converges to a
constant value. This means that (with respect to the laboratory frame
  $\Sigma_0$) the asymptotic mean square displacement of the free
  relativistic Brownian motions is again {\it normal}, i.e. it increases linearly with $t$. For
  completeness, we mention that according to our simulations the
  asymptotic relation \eqref{e:mean-a} holds in the relativistic case,
  too.
\par
In spite of these similarities between nonrelativistic and relativistic
theory, an essential difference consists in the explicit temperature
dependence of the limit value $6D^x$. As illustrated in Fig.  \ref{fig03} (b),
the numerical limit values $6D^x_{50}$, measured at time $t=50 \nu^{-1}$, are
reasonably well fitted by the  formula
\be\label{e:fitting}
D^x=\f{c^2}{\nu (\chi+6)},
\ee
which reduces to the nonrelativistic result \eqref{e:diffusion} in the limit
case $\chi\gg 6$ (low-temperature limit case). It remains as an open problem
for the future to find an analytic expression for the
relativistic diffusion constant $D^x$. Note also that, with the position being
the integral over the velocity degree of freedom, the different discretization
rules do not impact the asymptotic (long time)
result for the position diffusion coefficient.

\section{Summary}
\label{summary}

The challenge of this work has been  to extend our previous work
\cite{DuHa04a} on (1+1)-dimensional relativistic Brownian motions to
the (1+3)-dimensional case. To this end, we have introduced in
Sec. \ref{langevin_approach} a
(1+3)-dimensional relativistic generalization of the nonrelativistic
Langevin equations (LE). Analogous to the
nonrelativistic Ornstein-Uhlenbeck theory of Brownian motion
\cite{UhOr30,HaJu95,HaTo82,VK03}, it is implicitly assumed that the heat bath,
(which causes the stochastic motions of the particle) can  be regarded as an
isotropic, homogeneous fluid. Based on
this assumption, the relativistic equations of motions are
constructed such that they reduce  to the well-known nonrelativistic
LE in the limit case $c\to \infty$.
\par
In our relativistic version of the LE, the viscous friction
between Brownian particle and heat bath is modelled by a friction
tensor $\nu_{\ga\gb}$, exhibiting the same formal structure as
the pressure tensor of a perfect fluid \cite{Weinberg}. In particular,
this means that the friction tensor is uniquely determined by the value of the
(scalar) viscous friction coefficient $\nu$, as measured in  the
instantaneous rest frame of the particle. Similarly, the amplitude of
the stochastic force is also governed by a single parameter $D$,
specifying the Gaussian fluctuations of the heat bath, as seen in the
instantaneous rest frame of the particle.
\par
In Sec. \ref{lab} we have rewritten the relativistic LE in laboratory coordinates,
corresponding to a specific class of Lorentz
frames, in which the heat bath is assumed to be at rest at all
times. Further, it was shown that the relativistic equations can be
recast such that they contain ordinary \lq multiplicative\rq\space Gaussian
white noise. Analogous to nonrelativistic stochastic processes with \lq
multiplicative\rq\space noise, this leads to an ambiguity regarding
the interpretation of the stochastic differential equation; i.e.,
different discretization rules yield different Fokker-Planck
equations (Sec. \ref{FPE}).  Similar to the previous paper \cite{DuHa04a}, we
concentrated here on the three most popular discretization schemes,
corresponding to the pre-point rule proposed by Ito \cite{Ito44,Ito51}, the
mid-point rule by Stratonovich and Fisk \cite{St64,St66,Fisk63,Fisk65}, and the
post-point rule of Klimontovich and H\"anggi
\cite{Han78,Han80,Han82,Kl94}.  It was then shown in
Sec. \ref{FPE-r-lorentz}  that only the latter prescription,
i.e. the post-point discretization scheme, yields a
Fokker-Planck equation, whose stationary solution coincides with the
relativistic Maxwell-Boltzmann distribution, as e.g. known from the work of J\"uttner
\cite{Ju11}, Schay \cite{Sc61} and de Groot {\it et al.} \cite{DG80}.
\par
In Sec. \ref{numerics} we presented several numerical results (based
on the Ito scheme), including numerically obtained velocity
distribution functions and furthermore, the mean square displacement of free
Brownian particles. According to our findings, the relativistic mean square
displacement grows linearly with the laboratory coordinate time;
compared with nonrelativistic diffusions, however,  the temperature
dependence of the spatial diffusion constant becomes more intricate.
\par
Finally, we would like to emphasize that, so far, our approach of constructing
a relativistic Brownian motion dynamics is merely based on the condition
that -- given the (a priori prescribed)
stochastic force of the heat bath -- the relativistic LE has to converge to the well-known
nonrelativistic dynamical equation, if the Brownian
particle moves sufficiently slow.  In particular,  there presently
remains the open (and seemingly very difficult) problem of how to tackle
in a relativistically consistent manner the dynamics of all those
particles that constitute the heat bath (including their
coupling to the relativistic Brownian particle).

\begin{acknowledgments}
J. D. would like to thank S. Hilbert for valuable discussions and is
grateful to the Max-Planck-Gesellschaft for financial support.
\end{acknowledgments}


\bibliography{RBM,RelKin,StochCalc,FokkerPlanck}

\begin{thebibliography}{45}
\expandafter\ifx\csname natexlab\endcsname\relax\def\natexlab#1{#1}\fi
\expandafter\ifx\csname bibnamefont\endcsname\relax
  \def\bibnamefont#1{#1}\fi
\expandafter\ifx\csname bibfnamefont\endcsname\relax
  \def\bibfnamefont#1{#1}\fi
\expandafter\ifx\csname citenamefont\endcsname\relax
  \def\citenamefont#1{#1}\fi
\expandafter\ifx\csname url\endcsname\relax
  \def\url#1{\texttt{#1}}\fi
\expandafter\ifx\csname urlprefix\endcsname\relax\def\urlprefix{URL }\fi
\providecommand{\bibinfo}[2]{#2}
\providecommand{\eprint}[2][]{\url{#2}}

\bibitem[{\citenamefont{Einstein}(1905{\natexlab{a}})}]{Ei05c}
\bibinfo{author}{\bibfnamefont{A.}~\bibnamefont{Einstein}},
  \bibinfo{journal}{Ann. Phys. (Leipzig)} \textbf{\bibinfo{volume}{17}},
  \bibinfo{pages}{549} (\bibinfo{year}{1905}{\natexlab{a}}).

\bibitem[{\citenamefont{Einstein and von Smoluchowski}(1999)}]{EiSm}
\bibinfo{author}{\bibfnamefont{A.}~\bibnamefont{Einstein}} \bibnamefont{and}
  \bibinfo{author}{\bibfnamefont{M.}~\bibnamefont{von Smoluchowski}},
  \emph{\bibinfo{title}{Untersuchungen {\"u}ber die {T}heorie der {B}rownschen
  {B}ewegung/Abhandlungen {\"u}ber die {B}rownsche {B}ewegung und verwandte
  Erscheinungen}}, vol. \bibinfo{volume}{199} (\bibinfo{publisher}{Harri
  Deutsch}, \bibinfo{address}{Frankfurt}, \bibinfo{year}{1999}),
  \bibinfo{edition}{3rd} ed.

\bibitem[{\citenamefont{Uhlenbeck and Ornstein}(1930)}]{UhOr30}
\bibinfo{author}{\bibfnamefont{G.~E.} \bibnamefont{Uhlenbeck}}
  \bibnamefont{and} \bibinfo{author}{\bibfnamefont{L.~S.}
  \bibnamefont{Ornstein}}, \bibinfo{journal}{Phys. Rev.}
  \textbf{\bibinfo{volume}{36}}, \bibinfo{pages}{823} (\bibinfo{year}{1930}).

\bibitem[{\citenamefont{Chandrasekhar}(1943)}]{Ch43}
\bibinfo{author}{\bibfnamefont{S.}~\bibnamefont{Chandrasekhar}},
  \bibinfo{journal}{Rev. Mod. Phys.} \textbf{\bibinfo{volume}{15}},
  \bibinfo{pages}{1} (\bibinfo{year}{1943}).

\bibitem[{\citenamefont{Wang and Uhlenbeck}(1945)}]{WaUh45}
\bibinfo{author}{\bibfnamefont{M.~C.} \bibnamefont{Wang}} \bibnamefont{and}
  \bibinfo{author}{\bibfnamefont{G.~E.} \bibnamefont{Uhlenbeck}},
  \bibinfo{journal}{Rev. Mod. Phys.} \textbf{\bibinfo{volume}{17}},
  \bibinfo{pages}{323} (\bibinfo{year}{1945}).

\bibitem[{\citenamefont{Karatzas and Shreve}(1991)}]{KaSh91}
\bibinfo{author}{\bibfnamefont{I.}~\bibnamefont{Karatzas}} \bibnamefont{and}
  \bibinfo{author}{\bibfnamefont{S.~E.} \bibnamefont{Shreve}},
  \emph{\bibinfo{title}{Brownian Motion and Stochastic Calculus}}, no.
  \bibinfo{number}{113} in \bibinfo{series}{Graduate Texts in Mathematics}
  (\bibinfo{publisher}{Springer}, \bibinfo{address}{New York, Berlin},
  \bibinfo{year}{1991}), \bibinfo{edition}{2nd} ed.

\bibitem[{\citenamefont{Einstein}(1905{\natexlab{b}})}]{Ei05a}
\bibinfo{author}{\bibfnamefont{A.}~\bibnamefont{Einstein}},
  \bibinfo{journal}{Ann. Phys. (Leipzig)} \textbf{\bibinfo{volume}{17}},
  \bibinfo{pages}{891} (\bibinfo{year}{1905}{\natexlab{b}}).

\bibitem[{\citenamefont{Einstein}(1905{\natexlab{c}})}]{Ei05b}
\bibinfo{author}{\bibfnamefont{A.}~\bibnamefont{Einstein}},
  \bibinfo{journal}{Ann. Phys. (Leipzig)} \textbf{\bibinfo{volume}{18}},
  \bibinfo{pages}{639} (\bibinfo{year}{1905}{\natexlab{c}}).

\bibitem[{\citenamefont{Schay}(1961)}]{Sc61}
\bibinfo{author}{\bibfnamefont{G.}~\bibnamefont{Schay}}, Ph.D. thesis,
  \bibinfo{school}{Princeton University} (\bibinfo{year}{1961}),
  \bibinfo{note}{available through University Microfilms, Ann Arbor, Michigan,
  {\texttt https://wwwlib.umi.com}}.

\bibitem[{\citenamefont{Hakim}(1965)}]{Ha65}
\bibinfo{author}{\bibfnamefont{R.}~\bibnamefont{Hakim}}, \bibinfo{journal}{J.
  Math. Phys.} \textbf{\bibinfo{volume}{6}}, \bibinfo{pages}{1482}
  (\bibinfo{year}{1965}).

\bibitem[{\citenamefont{Dudley}(1965)}]{Du65}
\bibinfo{author}{\bibfnamefont{R.~M.} \bibnamefont{Dudley}},
  \bibinfo{journal}{Arkiv f\"or Matematik} \textbf{\bibinfo{volume}{6}},
  \bibinfo{pages}{241} (\bibinfo{year}{1965}).

\bibitem[{\citenamefont{Guerra and Ruggiero}(1978)}]{GuRu78}
\bibinfo{author}{\bibfnamefont{F.}~\bibnamefont{Guerra}} \bibnamefont{and}
  \bibinfo{author}{\bibfnamefont{P.}~\bibnamefont{Ruggiero}},
  \bibinfo{journal}{Lett. Nouvo Cimento} \textbf{\bibinfo{volume}{23}},
  \bibinfo{pages}{529} (\bibinfo{year}{1978}).

\bibitem[{\citenamefont{Boyer}(1979)}]{Bo79a}
\bibinfo{author}{\bibfnamefont{T.~H.} \bibnamefont{Boyer}},
  \bibinfo{journal}{Phys. Rev. D} \textbf{\bibinfo{volume}{19}},
  \bibinfo{pages}{1112} (\bibinfo{year}{1979}).

\bibitem[{\citenamefont{Ben-Ya'acov}(1981)}]{BY81}
\bibinfo{author}{\bibfnamefont{U.}~\bibnamefont{Ben-Ya'acov}},
  \bibinfo{journal}{Phys. Rev. D} \textbf{\bibinfo{volume}{23}},
  \bibinfo{pages}{1441} (\bibinfo{year}{1981}).

\bibitem[{\citenamefont{Morato and Viola}(1995)}]{MoVi95}
\bibinfo{author}{\bibfnamefont{L.~M.} \bibnamefont{Morato}} \bibnamefont{and}
  \bibinfo{author}{\bibfnamefont{L.}~\bibnamefont{Viola}}, \bibinfo{journal}{J.
  Math. Phys.} \textbf{\bibinfo{volume}{36}}, \bibinfo{pages}{4691}
  (\bibinfo{year}{1995}), \bibinfo{note}{erratum, {\it ibid.} {\bf 37}, 4769
  (1996)}.

\bibitem[{\citenamefont{Posilicano}(1997)}]{Po97}
\bibinfo{author}{\bibfnamefont{A.}~\bibnamefont{Posilicano}},
  \bibinfo{journal}{Lett. in Math. Phys.} \textbf{\bibinfo{volume}{42}},
  \bibinfo{pages}{85} (\bibinfo{year}{1997}).

\bibitem[{\citenamefont{Oron and Horwitz}(2003)}]{OrHo03}
\bibinfo{author}{\bibfnamefont{O.}~\bibnamefont{Oron}} \bibnamefont{and}
  \bibinfo{author}{\bibfnamefont{L.~P.} \bibnamefont{Horwitz}}
  (\bibinfo{year}{2003}), \bibinfo{note}{{\texttt arXiv:math-ph/0312003}}.

\bibitem[{\citenamefont{Franchi and {Le Jan}}(2004)}]{FrLJ04}
\bibinfo{author}{\bibfnamefont{J.}~\bibnamefont{Franchi}} \bibnamefont{and}
  \bibinfo{author}{\bibfnamefont{Y.}~\bibnamefont{{Le Jan}}}
  (\bibinfo{year}{2004}), \bibinfo{note}{{\texttt arXiv:math.PR/0403499}}.

\bibitem[{\citenamefont{Debbasch et~al.}(1997)\citenamefont{Debbasch, Mallick,
  and Rivet}}]{DeMaRi97}
\bibinfo{author}{\bibfnamefont{F.}~\bibnamefont{Debbasch}},
  \bibinfo{author}{\bibfnamefont{K.}~\bibnamefont{Mallick}}, \bibnamefont{and}
  \bibinfo{author}{\bibfnamefont{J.~P.} \bibnamefont{Rivet}},
  \bibinfo{journal}{J. Stat. Phys} \textbf{\bibinfo{volume}{88}},
  \bibinfo{pages}{945} (\bibinfo{year}{1997}).

\bibitem[{\citenamefont{Debbasch}(2004)}]{De04}
\bibinfo{author}{\bibfnamefont{F.}~\bibnamefont{Debbasch}},
  \bibinfo{journal}{J. Math. Phys.} \textbf{\bibinfo{volume}{45}},
  \bibinfo{pages}{2744} (\bibinfo{year}{2004}).

\bibitem[{\citenamefont{Stewart}(1971)}]{St71}
\bibinfo{author}{\bibfnamefont{J.~M.} \bibnamefont{Stewart}},
  \emph{\bibinfo{title}{Non-Equilibrium Relativistic Kinetic Theory}},
  vol.~\bibinfo{volume}{10} of \emph{\bibinfo{series}{Lecture Notes in
  Physics}} (\bibinfo{publisher}{Springer}, \bibinfo{address}{Berlin},
  \bibinfo{year}{1971}).

\bibitem[{\citenamefont{{de Groot} et~al.}(1980)\citenamefont{{de Groot}, {van
  Leeuwen}, and {van Weert}}}]{DG80}
\bibinfo{author}{\bibfnamefont{S.~R.} \bibnamefont{{de Groot}}},
  \bibinfo{author}{\bibfnamefont{W.}~\bibnamefont{{van Leeuwen}}},
  \bibnamefont{and} \bibinfo{author}{\bibfnamefont{C.~G.} \bibnamefont{{van
  Weert}}}, \emph{\bibinfo{title}{Relativistic Kinetic Theory: Principles and
  Applications}} (\bibinfo{publisher}{North-Holland},
  \bibinfo{address}{Amsterdam}, \bibinfo{year}{1980}).

\bibitem[{\citenamefont{Dunkel and H\"anggi}(2005)}]{DuHa04a}
\bibinfo{author}{\bibfnamefont{J.}~\bibnamefont{Dunkel}} \bibnamefont{and}
  \bibinfo{author}{\bibfnamefont{P.}~\bibnamefont{H\"anggi}},
  \bibinfo{journal}{Phys. Rev. E} \textbf{\bibinfo{volume}{71}},
  \bibinfo{pages}{016124} (\bibinfo{year}{2005}).

\bibitem[{\citenamefont{{Van Kampen}}(2003)}]{VK03}
\bibinfo{author}{\bibfnamefont{N.~G.} \bibnamefont{{Van Kampen}}},
  \emph{\bibinfo{title}{Stochastic Processes in Physics and Chemistry}}
  (\bibinfo{publisher}{North-Holland Personal Library},
  \bibinfo{address}{Amsterdam}, \bibinfo{year}{2003}).

\bibitem[{\citenamefont{H\"anggi and Thomas}(1982{\natexlab{a}})}]{HaTo82}
\bibinfo{author}{\bibfnamefont{P.}~\bibnamefont{H\"anggi}} \bibnamefont{and}
  \bibinfo{author}{\bibfnamefont{H.}~\bibnamefont{Thomas}},
  \bibinfo{journal}{Phys. Rep.} \textbf{\bibinfo{volume}{88}},
  \bibinfo{pages}{207} (\bibinfo{year}{1982}{\natexlab{a}}).

\bibitem[{\citenamefont{Ito}(1944)}]{Ito44}
\bibinfo{author}{\bibfnamefont{K.}~\bibnamefont{Ito}}, \bibinfo{journal}{Proc.
  Imp. Acad. Tokyo} \textbf{\bibinfo{volume}{20}}, \bibinfo{pages}{519}
  (\bibinfo{year}{1944}).

\bibitem[{\citenamefont{Ito}(1951)}]{Ito51}
\bibinfo{author}{\bibfnamefont{K.}~\bibnamefont{Ito}}, \bibinfo{journal}{Mem.
  Amer. Mathem. Soc.} \textbf{\bibinfo{volume}{4}}, \bibinfo{pages}{51}
  (\bibinfo{year}{1951}).

\bibitem[{\citenamefont{Stratonovich}(1964)}]{St64}
\bibinfo{author}{\bibfnamefont{R.~L.} \bibnamefont{Stratonovich}},
  \bibinfo{journal}{Vestnik Moskov. Univ., Ser. I: Mat., Mekh.}
  \textbf{\bibinfo{volume}{1}}, \bibinfo{pages}{3} (\bibinfo{year}{1964}).

\bibitem[{\citenamefont{Stratonovich}(1966)}]{St66}
\bibinfo{author}{\bibfnamefont{R.~L.} \bibnamefont{Stratonovich}},
  \bibinfo{journal}{J. SIAM Control Optim.} \textbf{\bibinfo{volume}{4}},
  \bibinfo{pages}{362} (\bibinfo{year}{1966}).

\bibitem[{\citenamefont{Fisk}(1963)}]{Fisk63}
\bibinfo{author}{\bibfnamefont{D.}~\bibnamefont{Fisk}}, \bibinfo{journal}{Ph.D
  thesis, Michigan State University, Dept. of Statistics}
  (\bibinfo{year}{1963}).

\bibitem[{\citenamefont{Fisk}(1965)}]{Fisk65}
\bibinfo{author}{\bibfnamefont{D.}~\bibnamefont{Fisk}},
  \bibinfo{journal}{Trans. Amer. Math. Soc.} \textbf{\bibinfo{volume}{120}},
  \bibinfo{pages}{369} (\bibinfo{year}{1965}).

\bibitem[{\citenamefont{H\"anggi}(1978)}]{Han78}
\bibinfo{author}{\bibfnamefont{P.}~\bibnamefont{H\"anggi}},
  \bibinfo{journal}{Helv. Phys. Acta} \textbf{\bibinfo{volume}{51}},
  \bibinfo{pages}{183} (\bibinfo{year}{1978}).

\bibitem[{\citenamefont{H\"anggi}(1980)}]{Han80}
\bibinfo{author}{\bibfnamefont{P.}~\bibnamefont{H\"anggi}},
  \bibinfo{journal}{Helv. Phys. Acta} \textbf{\bibinfo{volume}{53}},
  \bibinfo{pages}{491} (\bibinfo{year}{1980}).

\bibitem[{\citenamefont{H\"anggi and Thomas}(1982{\natexlab{b}})}]{Han82}
\bibinfo{author}{\bibfnamefont{P.}~\bibnamefont{H\"anggi}} \bibnamefont{and}
  \bibinfo{author}{\bibfnamefont{H.}~\bibnamefont{Thomas}},
  \bibinfo{journal}{Phys. Rep.} \textbf{\bibinfo{volume}{88}},
  \bibinfo{pages}{207} (\bibinfo{year}{1982}{\natexlab{b}}), \bibinfo{note}{see
  pp. 292--294}.

\bibitem[{\citenamefont{Klimontovich}(1994)}]{Kl94}
\bibinfo{author}{\bibfnamefont{Y.~L.} \bibnamefont{Klimontovich}},
  \bibinfo{journal}{Physics-Uspekhi} \textbf{\bibinfo{volume}{37}},
  \bibinfo{pages}{737} (\bibinfo{year}{1994}).

\bibitem[{\citenamefont{J\"uttner}(1911)}]{Ju11}
\bibinfo{author}{\bibfnamefont{F.}~\bibnamefont{J\"uttner}},
  \bibinfo{journal}{Ann. Phys. (Leipzig)} \textbf{\bibinfo{volume}{34}},
  \bibinfo{pages}{856} (\bibinfo{year}{1911}).

\bibitem[{\citenamefont{Synge}(1957)}]{Sy57}
\bibinfo{author}{\bibfnamefont{J.~L.} \bibnamefont{Synge}},
  \emph{\bibinfo{title}{The Relativistic Gas}}
  (\bibinfo{publisher}{North-Holland}, \bibinfo{address}{Amsterdam},
  \bibinfo{year}{1957}).

\bibitem[{\citenamefont{Weinberg}(1972)}]{Weinberg}
\bibinfo{author}{\bibfnamefont{S.}~\bibnamefont{Weinberg}},
  \emph{\bibinfo{title}{Gravitation and Cosmology}} (\bibinfo{publisher}{John
  Wiley \& Sons}, \bibinfo{year}{1972}).

\bibitem[{\citenamefont{Goldstein}(1980)}]{Go80}
\bibinfo{author}{\bibfnamefont{H.}~\bibnamefont{Goldstein}},
  \emph{\bibinfo{title}{Classical Mechanics}}, Addison-Wesley series in physics
  (\bibinfo{publisher}{Addison-Wesley}, \bibinfo{address}{Reading},
  \bibinfo{year}{1980}), \bibinfo{edition}{2nd} ed.

\bibitem[{\citenamefont{Schwabl}(2000)}]{Schwabl}
\bibinfo{author}{\bibfnamefont{F.}~\bibnamefont{Schwabl}},
  \emph{\bibinfo{title}{Statistische Mechanik}} (\bibinfo{publisher}{Springer},
  \bibinfo{address}{Berlin, Heidelberg}, \bibinfo{year}{2000}).

\bibitem[{\citenamefont{Zeidler}(1996)}]{Te1}
\bibinfo{editor}{\bibfnamefont{E.}~\bibnamefont{Zeidler}}, ed.,
  \emph{\bibinfo{title}{Teubner-Taschenbuch der Mathematik}},
  vol.~\bibinfo{volume}{1} (\bibinfo{publisher}{B. G. Teubner},
  \bibinfo{address}{Stuttgart, Leipzig}, \bibinfo{year}{1996}),
  \bibinfo{edition}{19th} ed., \bibinfo{note}{p. 1104}.

\bibitem[{\citenamefont{Liboff}(1990)}]{Liboff90}
\bibinfo{author}{\bibfnamefont{R.~L.} \bibnamefont{Liboff}},
  \emph{\bibinfo{title}{Kinetic Theory}} (\bibinfo{publisher}{Prentice Hall,
  Inc.}, \bibinfo{address}{Englewood Cliffs, New Jersey 07632},
  \bibinfo{year}{1990}).

\bibitem[{\citenamefont{Misner et~al.}(2000)\citenamefont{Misner, Thorne, and
  Wheeler}}]{MiThWe00}
\bibinfo{author}{\bibfnamefont{C.~W.} \bibnamefont{Misner}},
  \bibinfo{author}{\bibfnamefont{K.~S.} \bibnamefont{Thorne}},
  \bibnamefont{and} \bibinfo{author}{\bibfnamefont{J.~A.}
  \bibnamefont{Wheeler}}, \emph{\bibinfo{title}{Gravitation}}
  (\bibinfo{publisher}{W. H. Freeman and Co.}, \bibinfo{address}{New York},
  \bibinfo{year}{2000}), \bibinfo{note}{23rd printing}.

\bibitem[{Ma0(2003)}]{Ma03}
\emph{\bibinfo{title}{Mathematica 5.0.1.0}}, \bibinfo{organization}{Wolfram
  Research, Inc.} (\bibinfo{year}{2003}).

\bibitem[{\citenamefont{H\"anggi and Jung}(1995)}]{HaJu95}
\bibinfo{author}{\bibfnamefont{P.}~\bibnamefont{H\"anggi}} \bibnamefont{and}
  \bibinfo{author}{\bibfnamefont{P.}~\bibnamefont{Jung}},
  \bibinfo{journal}{Adv. Chem. Phys.} \textbf{\bibinfo{volume}{89}},
  \bibinfo{pages}{239} (\bibinfo{year}{1995}).

\end{thebibliography}

\appendix

\section{Stationary solutions of relativistic Fokker-Planck equations}
\label{a:solution}

We seek the stationary solutions $f(\bs p)$ of the FPE \eqref{e:continuity},
which lead to vanishing currents
\be\label{e:j=0}
j^i_\mrm{I/SF/HK}(\bs p)\equiv 0.
\ee
We use the following Ansatz
\be
f(\bs p)=C\gc^{-\ga}\exp(-\chi \gc),
\ee
where $C>0$ is a normalization constant and
\be
\gc=\left(1+\f{p_i p^i}{m^2c^2}\right)^{1/2}.
\ee
The parameters $\ga$ and $\chi$ have to be determined from the condition \eqref{e:j=0}.
Therefore we have the following partial derivatives
\bse
\be
\f{\p \gc}{\p p^j}&=&\f{p_j}{\gc m^2c^2},\\
\f{\p f}{\p p^j}&=&\f{-p_j}{\gc m^2c^2}\left(\f{\ga}{\gc}+\chi\right)f,
\ee
and furthermore the divergence
\be\label{e:divergence_A}
\f{\p}{\p p^j} (A^{-1})^{ij}
=\f{\p}{\p p^j}\left[\left(\gd^{ij}+\f{ p^ip^j}{m^2c^2}\right)\f{1}{\gc}\right]
=\f{3 p^i}{\gc m^2c^2}.
\ee
\ese

\subsection{Ito current}
\label{a:ito}

For the Ito current $\bs j_\mrm{I}$ from \eqref{e:ito_current}, the condition \eqref{e:j=0} yields
\be
0
&\equiv& \notag
-j_\mrm{I}^i(\bs p)\\
&=& \notag
\nu p^i f+ D f \f{\p}{\p p^j} (A^{-1})^{ij}
+ D \, (A^{-1})^{ij}  \f{\p f}{\p p^j} \\
&=&\notag
\nu p^i f+ D f \f{3 p^i}{\gc m^2c^2}+
D \left(\gd^{ij}+\f{ p^ip^j}{m^2c^2}\right)
\f{-p_j}{\gc^2 m^2c^2}\left(\f{\ga}{\gc}+\chi\right)f\\
&=&
p^i f\left[
\nu+ \f{3 D}{\gc m^2c^2}- \f{D}{m^2c^2}\left(\f{\ga}{\gc}+\chi\right)
\right],
\ee
which is fulfilled for $\ga=3$ and  $\chi=\nu m^2c^2/D$.

\subsection{Stratonovich-Fisk current}
\label{a:SF}

For the Stratonovich-Fisk current $\bs j_\mrm{SF}$ from  \eqref{e:stratonovich_current}, the condition \eqref{e:j=0} becomes
\be
0
&\equiv&\notag
-j^i_\mrm{SF}(\bs p)\\
&=&\notag
\nu p^i f+D\, {(L^{-1})^i}_k
\f{\p}{\p p_j}{\left[(L^{-1})^\top\right]^k}_j  \; f\\
&=&\label{e:j=0_SF}
\nu p^i f +
D\, {(L^{-1})^i}_k\;f \f{\p}{\p p_j}{\left[(L^{-1})^\top\right]^k}_j +
D \, {(A^{-1})^i}_j  \f{\p f}{\p p_j},
\ee
where $\bs L(\bs p)^{-1}$ has been given in \eqref{e:inverse_L}. A
lengthy, though straightforward calculation shows that
\be
{(L^{-1})^i}_k
\f{\p}{\p p_j}
{\left[(L^{-1})^\top\right]^k}_j=\f{3 p^i}{2\gc m^2c^2}.
\ee
Inserting this into \eqref{e:j=0_SF}, one finds
\be
0&=&\notag
\nu p^i f+ D f \f{3 p^i}{2\gc m^2c^2}+
D \left(\gd^{ij}+\f{ p^ip^j}{m^2c^2}\right)
\f{-p_j}{\gc^2 m^2c^2}\left(\f{\ga}{\gc}+\chi\right)f\\
&=&p^i f\left[
\nu+ \f{3 D}{2\gc m^2c^2}- \f{D}{m^2c^2}\left(\f{\ga}{\gc}+\chi\right)
\right],
\ee
which is fulfilled for $\ga=3/2$ and  $\chi=\nu m^2c^2/D$.

\subsection{HK-current}
\label{a:HK}

For the HK-current $\bs j_\mrm{K}$ from \eqref{e:K_current}, the condition \eqref{e:j=0} yields
\be
0
&\equiv& \notag
-j_\mrm{HK}^i(\bs p)\\
&=& \notag
\nu p^i f+ D \, (A^{-1})^{ij}  \f{\p f}{\p p^j} \\
&=&\notag
\nu p^i f+ D \left(\gd^{ij}+\f{ p^ip^j}{m^2c^2}\right)
\f{-p_j}{\gc^2 m^2c^2}\left(\f{\ga}{\gc}+\chi\right)f\\
&=&
p^i f\left[\nu- \f{D}{m^2c^2}\left(\f{\ga}{\gc}+\chi\right)
\right],
\ee
which is fulfilled for $\ga=0$ and  $\chi=\nu m^2c^2/D$.

\end{document}